\documentclass[12pt]{article}

\usepackage{epsfig,amsmath,amssymb,verbatim,mathrsfs,hyperref}
\usepackage{xspace}

\usepackage{slashed}

\def\beq{\begin{eqnarray}}
\def\eeq{\end{eqnarray}}
\def\bea{\begin{eqnarray}}
\def\eea{\end{eqnarray}}

\def\tev{\, {\rm TeV}}
\def\gev{\, {\rm GeV}}
\def\mev{\, {\rm MeV}}

\newcommand{\gsim}{\lower.7ex\hbox{$\;\stackrel{\textstyle>}{\sim}\;$}}
\newcommand{\lsim}{\lower.7ex\hbox{$\;\stackrel{\textstyle<}{\sim}\;$}}

\newcommand{\hc}{\;\mathrm{h.c.}}

\newcommand{\nnmb}{\nonumber}
\newcommand{\del}{\partial}

\newcommand{\lrf}[2]{\left(\frac{#1}{#2}\right)}

\newcommand{\lag}{\mathscr{L}}

\newcommand{\kev}{\mathrm{keV}}

\newcommand{\GeV}{\mathrm{GeV}}

\newcommand{\Tfo}{T_\mathrm{fo}}
\newcommand{\micromegas}{\texttt{micrOMEGAs}\xspace}

\begin{document}
\begin{titlepage}
\noindent
SLAC-PUB-16419
\vspace{1cm}
\begin{center}
  \begin{Large}
    \begin{bf}
Compressing the Inert Doublet Model
     \end{bf}
  \end{Large}
\end{center}
\vspace{0.2cm}
\begin{center}
\begin{large}
Nikita Blinov$^{(a,b,c)}$, Jonathan Kozaczuk$^{(a)}$,\\ 
David E. Morrissey$^{(a)}$, and Alejandro de la Puente$^{(a,d)}$
\end{large}
\vspace{1cm}\\
\begin{it}
(a) TRIUMF, 4004 Wesbrook Mall, Vancouver, BC V6T 2A3, Canada\vspace{0.2cm}\\
(b) Department of Physics and Astronomy, University of British Columbia, Vancouver, BC V6T 1Z1, Canada\vspace{0.2cm}\\
(c) SLAC National Accelerator Laboratory, 2575 Sand Hill Road, Menlo Park, CA, 94025, USA \vspace{0.2cm}\\
(d) Department of Physics, Carleton University, Ottawa, ON K1S 5B6, Canada
\vspace{0.5cm}\\
email: \emph{\texttt{nblinov@slac.stanford.edu}}, \emph{\texttt{jkozaczuk@triumf.ca}},\\
\emph{\texttt{dmorri@triumf.ca}}, \emph{\texttt{apuente@physics.carleton.ca}}
\vspace{0.2cm}
\end{it}
\end{center}
\center{\today}

\begin{abstract}

The Inert Doublet Model relies on a discrete symmetry to prevent couplings of the new scalars to Standard Model fermions. This stabilizes the lightest inert state, which can then contribute to the observed dark matter density. In the presence of additional approximate symmetries, the resulting spectrum of exotic scalars can be compressed. Here, we study the phenomenological and cosmological implications of this scenario. We derive new limits on the compressed Inert Doublet Model from LEP, and outline the prospects for exclusion and discovery of this model at dark matter experiments, the LHC, and future colliders.

\end{abstract}

\end{titlepage}

\setcounter{page}{2}

%%%%%%%%%%%%%%%%%%%%%%%%%%%%%%%%%%%%%%%%%%%%%%%%%%%%%%%%%%%%%%%%%%%%%%

\section{Introduction\label{sec:intro}}

The Inert Doublet Model~(IDM) extends the Standard Model~(SM) 
with a scalar $SU(2)_L$ doublet that is odd under an unbroken
$\mathbb{Z}_2$ symmetry~\cite{Deshpande:1977rw} .  
This symmetry forbids direct couplings of the new doublet
to the SM fermions while allowing couplings to the Higgs, 
and implies that the lightest state derived from 
the new doublet is stable.  For these reasons, the IDM has been 
studied extensively as a model of dark 
matter~(DM)~\cite{Barbieri:2006dq,LopezHonorez:2006gr}, 
as a simple module to investigate deviations in the properties of 
the Higgs boson~\cite{Barbieri:2006dq,Cao:2007rm}, 
and as a general source of new missing-energy signals 
in collider experiments~\cite{Cao:2007rm,Dolle:2009ft}. 
Many other scenarios that postulate additional weakly-charged 
scalars can also be mapped onto (one or many copies of) the IDM~\cite{Hirsch:2010ru,Boucenna:2011tj,Brown:2010ke,Lavoura:2012cv,Martin:2013fta,Kephart:2015oaa,Nagata:2015dma,Carmona:2015haa},
and the model can help to induce a strongly first-order electroweak
phase transition suitable for electroweak baryogenesis~\cite{Chowdhury:2011ga,Borah:2012pu,Gil:2012ya,AbdusSalam:2013eya,Blinov:2015sna,Blinov:2015vma}.

The scalar sector of the IDM contains two doublets $H_{1,2}$ transforming under 
$SU(3)_c\times SU(2)_L\times U(1)_Y$ as  $(\mathbf{1},\mathbf{2},1/2)$ 
with a tree-level potential given by
\beq
V &=& \mu_1^2|H_1|^2 + \mu_2^2|H_2|^2 + \lambda_1|H_1|^4 + \lambda_2|H_2|^4
\label{eq:vidm}\\
&&~~+\lambda_3|H_1|^2|H_2|^2 + \lambda_4|H_1^{\dagger}H_2|^2
+ \frac{\lambda_5}{2}[(H_1^{\dagger}H_2)^2 + \hc] \ .
\nnmb
\eeq
We identify $H_1$ with the SM Higgs field and assume that
only it gets a vacuum expectation value~(VEV)
with $\langle H_1\rangle = v \simeq 174\,\gev$.
The tree-level scalar mass spectrum is then
\beq
m_h^2 &=& -2\mu_1^2~=~{4\lambda_1v^2}
\label{eq:mhiggs}\\
m_H^2 &=& \mu_2^2 + (\lambda_3+\lambda_4+\lambda_5)v^2
\label{eq:mhh}\\
m_A^2 &=& \mu_2^2 + (\lambda_3+\lambda_4-\lambda_5)v^2
\label{eq:mha}\\
m_{H^\pm}^2\!\!\! &=& \mu_2^2 + \lambda_3v^2 \ . 
\label{eq:mhpm}
\eeq
The SM-like Higgs boson $h$ comes entirely from $H_1$ while the rest
come from $H_2$.  The other components of $H_1$ are eaten 
by the weak vector bosons. 

It is straightforward to check that Eq.~\eqref{eq:vidm} is the most 
general dimension-four scalar potential consistent with 
the discrete $\mathbb{Z}_2$
under which $H_1\to H_1$ and $H_2\to -H_2$~\cite{Hambye:2009pw}.  
This symmetry also forbids $H_2$ from coupling directly to SM fermions.  
The form of the scalar potential is constrained further in the 
presence of additional approximate symmetries.
Extending the $\mathbb{Z}_2$ to a global $U(1)$ acting only on $H_2$
forces $\lambda_5=0$ and leads to $m_H = m_A$~\cite{Barbieri:2006dq}, 
while extending the $\mathbb{Z}_2$ to a global $SU(2)$ acting on $H_2$ 
forces both $\lambda_4=\lambda_5=0$ and makes all three exotic scalars 
degenerate~\cite{Gerard:2007kn}.
These additional symmetries could be the parent symmetry of the stabilizing 
$\mathbb{Z}_2$, or they could arise from some other source. 
Motivated by these possible extended symmetries,
we investigate the IDM in the limit of parametrically small
$|\lambda_4|$ and $|\lambda_5|$.  The resulting spectrum of exotic
scalars is then compressed, which has significant phenomenological implications.

In this paper, we study the IDM in the compressed limit, 
focusing on the properties of IDM dark matter in this regime and the prospects
for discovering the new scalars using existing and future collider data.
Beginning in Section~\ref{sec:split} we 
discuss the extent to which imposing additional approximate symmetries can 
produce a compressed mass spectrum in the IDM.  
Next, in Section~\ref{sec:basic} we review basic existing constraints
on the IDM.  In Section~\ref{sec:cosmo}, we study the dark matter and
cosmological limits on the theory with nearly mass-degenerate scalars.
Sections~\ref{sec:lep},~\ref{sec:lhc}, and~\ref{sec:fc} investigate 
the current collider limits and future discovery prospects of 
the compressed IDM at LEP, the LHC, and proposed future colliders.
Finally, Section~\ref{sec:conc} is reserved for our conclusions.

\section{Small Mass Splittings\label{sec:split}}

  A compressed inert scalar spectrum arises for values of 
$|\lambda_4|$ and $|\lambda_5|$ much smaller than unity.
From Eqs.~(\ref{eq:mhh},\ref{eq:mha},\ref{eq:mhpm}) 
the mass splittings in this limit are
\beq
\Delta^0 &=& m_A-m_H ~~\hspace{1mm}\simeq~~ -\frac{\lambda_5v^2}{m_{H}} \ , \\
\Delta^{\pm} &=& m_{H^{\pm}}-m_H ~\simeq~ 
-\frac{(\lambda_4+\lambda_5)v^2}{2m_{H}}  \ .
\eeq
Small values of $|\lambda_4|$ and $|\lambda_5|$ can arise in
a technically natural way if the theory has additional
approximate symmetries. The three relevant symmetries in this regard are 
a global $U(1)$ acting on $H_2$, a global $SU(2)$ acting 
on $H_2$, and an extension of the custodial symmetry of the SM 
to include the inert doublet.  We consider all three here, 
and investigate their implications for the mass spectrum of the 
new scalars and the lifetimes of the heavier states.
Our primary conclusion is that the neutral mass splitting $\Delta^0$
can be arbitrarily small, while the charged mass splitting $\Delta^\pm$
tends to be larger than about a GeV.

\subsection{Symmetries and Splittings}

  For $\lambda_5=0$ in the potential, Eq.~\eqref{eq:vidm}, the theory
has an enhanced symmetry under global $U(1)_2$ transformations acting
on $H_2$ alone~\cite{Barbieri:2006dq}.  
This leads to $m_H=m_A$, and it is convenient in this case
to assemble these two real scalars into a single neutral complex scalar.  
Note as well that $U(1)_2$ is trivially non-anomalous and 
its generator commutes with all the gauge generators; thus it can be
an exact symmetry of the theory.

  With $\lambda_4=0=\lambda_5$, the scalar potential is also
invariant under global $SU(2)_2$ transformations acting on $H_2$ alone.
This leads to $m_H=m_A = m_{H^{\pm}}$ at tree-level.  Both $\lambda_4$ and
$\lambda_5$ can also be set to zero by extending the custodial symmetry
of the SM to include the new doublet~\cite{Gerard:2007kn}.\footnote{
There is a one-parameter family of ways to extend the custodial symmetry.  
General choices of the embedding parameter force $\lambda_4=0=\lambda_5$, 
although special choices also allow 
$\lambda_4=\pm\lambda_5$~\cite{Gerard:2007kn}.}
However, in contrast to the global $U(1)_2$ described above, 
both of these symmetries of the scalar potential
are broken explicitly by gauging $SU(2)_L\times U(1)_Y$.
Specifically, the $SU(2)_2$ generator does not commute with 
those of $SU(2)_L$, while $U(1)_Y$ breaks the extended custodial invariance
just like in the SM.  

  Based on these considerations, it is technically natural for $|\lambda_5|$
to be arbitrarily small, while very suppressed values of $|\lambda_4|$ 
may require some degree of fine tuning.  This can be seen explicitly in the 
one-loop renormalization group~(RG) equations 
of these couplings~\cite{Goudelis:2013uca}:
\beq
(4\pi)^2\frac{d\lambda_4}{dt} &=& - 3\lambda_4(3g^2+g^{\prime 2})
+ 4\lambda_4(\lambda_1+\lambda_2+2\lambda_3+\lambda_4)
+2\lambda_4(3y_t^2+3y_b^2+y_{\tau}^2)
\label{eq:l4run}\\ 
&&+~3g^2g^{\prime 2} ~+~8\lambda_5^2\nnmb\\
(4\pi)^2\frac{d\lambda_5}{dt} &=& - 3\lambda_5(3g^2+g^{\prime 2})
+ 4\lambda_5(\lambda_1+\lambda_2+2\lambda_3+3\lambda_4)
+2\lambda_5(3y_t^2+3y_b^2+y_{\tau}^2) \ ,
\label{eq:l5run}
\eeq
where $t = \ln(\mu/\mu_0)$, and $\mu$ is the renormalization scale
with $\mu_0=100\,\gev$.
As expected, if $\lambda_5$ vanishes at any reference scale it
will vanish at any other.  In contrast, even if $\lambda_4=0 = \lambda_5$ 
at some input scale $M_{in}$, it will 
be regenerated by the $g^2g^{\prime 2}$ term in the second line 
of Eq.~\eqref{eq:l4run}. The appearance of
both $g$ and $g^{\prime}$ in this contribution coincides with the explicit
breaking of both the $SU(2)_2$ and extended custodial symmetries
by electroweak gauging.  

\begin{figure}[ttt]
 \begin{center}
         \includegraphics[width = 0.65\textwidth]{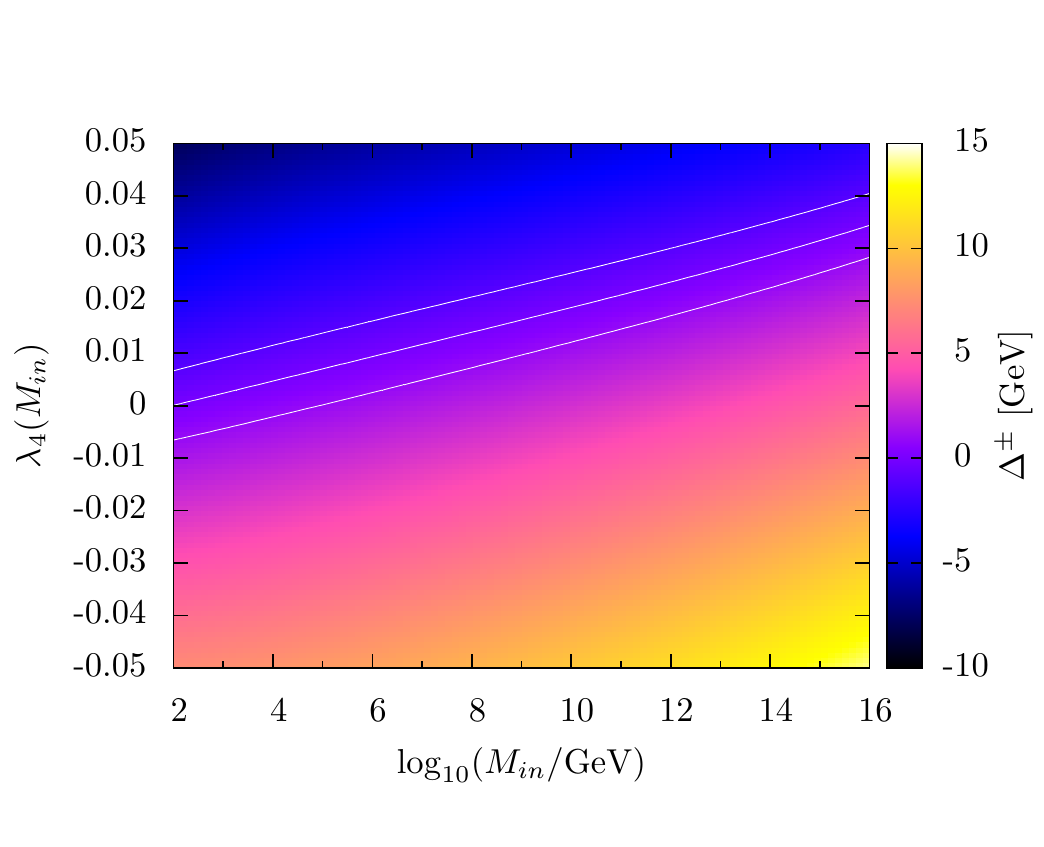}
 \end{center}
\vspace{-1cm}
 \caption{Tree-level charged-neutral mass splittings $\Delta^{\pm}$ 
for a given value of $\lambda_4(M_{in})$ specified at the input scale $M_{in}$
with $m_H=100\,\gev$,
$\lambda_2(m_Z)=0.1$, $\lambda_3(m_Z)=0.05$, 
and $\lambda_5 = 0$.  The solid white lines show contours of
$\Delta^{\pm} = (0\pm 1)\,\gev$.}
 \label{fig:delm}
 \end{figure}

  To examine the tree-level mass splitting induced by 
the RG running of $\lambda_4$, we compute the low-scale mass difference 
$\Delta^{\pm}$ between $H^{\pm}$ and $H^0$ assuming small initial 
values of $\lambda_4(M_{in})$ defined at the high input scale $M_{in}$.  
The RG evolution is implemented using the equations for all 
the relevant couplings tabulated in Ref.~\cite{Goudelis:2013uca}.  
In doing so, we fix the gauge and Yukawa couplings and $\lambda_1$
according to their measured values, and we set $m_H=100\,\gev$,
$\lambda_2(m_Z)=0.1$, $\lambda_3(m_Z)=0.05$, 
and $\lambda_5 = 0$. The results are 
mostly independent of these choices provided they are much smaller
than the electroweak gauge couplings. 
The resulting RG-induced low-scale mass splitting $\Delta^{\pm}$ 
is shown in Fig.~\ref{fig:delm}, with the solid lines indicating
where $\Delta^{\pm} = (0\pm1)\,\gev$.
We find typical mass splittings on the order of a few GeV
for vanishing initial values of $\lambda_4(M_{in})$ and $\lambda_5(M_{in})$
unless $M_{in}$ is very close to the weak scale.

  In addition to the RG-induced mass splitting, there is also a finite
contribution to $\Delta^{\pm}$ from electroweak symmetry breaking.  
At one-loop order it is~\cite{Cirelli:2005uq}
\beq
\Delta^{\pm}_{\text{one-loop}} ~=~ 
\frac{\alpha_Ws_W^2}{4\pi}m_{H^{\pm}}\,f(m_Z/m_{H^{\pm}}) \ , 
\eeq
where the function $f(x)$ is given by
\beq
f(x) = -\frac{x}{4}\left(
2x^3\ln x 
+(x^2-4)^{3/2}\ln\left[(x^2-2-x\sqrt{x^2-4})/2\right]
\right) \ .
\eeq
Note that we have already accounted for the UV-divergent part listed
in Ref.~\cite{Cirelli:2005uq} by the RG analysis above.
We find that this additional splitting is less than about 
$\Delta^{\pm} \lesssim 100\,\mev$ for the parameter regions of greatest
interest from the standpoint of collider searches, $m_H\lesssim 100\,\gev$,
and is almost always subdominant to the RG-induced
tree-level splitting considered above.

Based on these results, in the remainder of this paper we will study
the IDM in the compressed limit with arbitrarily small mass splittings 
$\Delta^0$ between the neutral $H$ and $A$ states, but we will assume
somewhat larger mass splittings among these and the charged state,
with $|\Delta^{\pm}| \gtrsim 1\,\gev$.  Reducing $\Delta^{\pm}$
much below this typically requires either a fortuitous accident or a significant
fine-tuning for $M_{in} \gg m_Z$.  We will also assume $\Delta^0 \geq 0$
with no loss of generality, as well as $\Delta^{\pm} > 0$
to avoid a stable charged relic.

\subsection{Decay Lifetimes}

A compressed mass spectrum in the IDM can suppress
the decay rates of the heavier states down to the lightest.
As mentioned above, we will consider arbitrarily small neutral
splittings $\Delta^0$ but we will focus on charged mass splittings
above $\Delta^\pm \gtrsim 1\,\gev$.  These values generally cause the
$H^\pm$ decay promptly on collider time scales but allow the heavier
neutral state $A$ to be long-lived on both collider and cosmological time scales.

  The dominant decay channels for both $A$ and $H^\pm$ are 
$A\to HZ^*$ and $H^\pm\to HW^{\pm*}$, with off-shell weak vector bosons.  
When $\Delta^{0,\pm}\ll m_W$, the vector bosons can be integrated out to
give the leading effective interactions
\beq
-\lag_{\mathrm{eff}} &\supset& \frac{g^2}{2m_W^2}\left[
A\stackrel{\leftrightarrow}{\del_{\mu}}\!H\sum_i\bar{f}_i\gamma^{\mu}
\left(a_V^i+a_A^i\gamma^5\right)f_i\right.
\label{eq:leff}
\\
&&
\left.
~~~~~+~~iH^+\!\stackrel{\leftrightarrow}{\del_{\mu}}\!(H-iA)
\sum_{jk}\bar{f}_j\gamma^{\mu}
\left(c_V^{\;jk}+c_A^{\;jk}\gamma^5\right)f_k^\prime
+ (\text{h.c.})\right] \ ,\nnmb
\eeq
where the couplings are given by
\beq
a_v^i = \frac{1}{2}(t^3_i-2Q_is^2_W),~~~~~a_A^i = -\frac{1}{2}t^3_i \ ,
\eeq
with $i$ running over all SM fermion species, and 
\beq
c_V^{\;jk} ~=~ -c_A^{\;jk} &=& \frac{1}{2\sqrt{2}}\delta^{jk}~~~~~~~~(\text{leptons})\\
&=& \frac{1}{2\sqrt{2}}V^{jk}_{\rm CKM}~~~~(\text{quarks})
\eeq
where $j$ runs over up-type fermions and $k$ over down-type,
and $V_{\rm CKM}$ is the Cabbibo-Kobayashi-Maskawa matrix.

  The decay width for $A\to H+f\bar{f}$ derived from these interactions is
approximately
\beq
\Gamma(A\to H) ~=~ \frac{1}{120\pi^3}\frac{g^4}{m_W^4}(\Delta^{0})^5\,
\sum_iN_c^i\left[(a_V^i)^2+(a_A^i)^2\right]\times\Theta(\Delta^0-2m_i)
\label{eq:gama}
\eeq
with $N_c^i$ the number of colours of the $i$-th species.
Similarly, the width for the charged $H^+\to H+f\bar{f}^\prime$ channels is
\beq
\Gamma(H^+\to H) ~=~ \frac{1}{120\pi^3}\frac{g^4}{m_W^4}(\Delta^{\pm})^5\,
\sum_{jk}N_c^{j}\left[|c_V^{\;jk}|^2+|c_A^{\;jk}|^2\right]
\times\Theta(\Delta^\pm-m_j-m_k)
\ .
\label{eq:gamhp}
\eeq
A similar expression applies for $H^+\to Af\bar{f}^\prime$ decays.

These expressions are only sensible when the hadronic final states can
be reliably treated as partons.  In our numerical estimates, we apply them
down to $\Delta=2\,\gev$.  For smaller splittings,
the hadronic decays can be handled similarly to tau leptons~\cite{Kuhn:1990ad}
or nearly-degenerate electroweakino 
superpartners~\cite{Chen:1995yu,Chen:1996ap,Chen:1999yf}.  No hadronic
modes are available for $\Delta < m_{\pi}$, while the one-pion
mode $A\to H\pi^0$ ($H^+\to H\pi^+$) is expected to dominate 
the hadronic width for $m_{\pi} < \Delta \lesssim 1\,\gev$~\cite{Chen:1999yf}.
Thus, to estimate the hadronic decay width for $m_{\pi} < \Delta < 2\,\gev$
we compute the one-pion width explicitly up to 1 GeV and then interpolate the
result smoothly to the partonic result at 2~GeV.  A more detailed treatment
following Refs.~\cite{Kuhn:1990ad,Chen:1999yf} could be applied 
if specific partial widths are desired.

  To compute the one-pion decay widths, we match the quark-level
operators of Eq.~\eqref{eq:leff} to the axial isospin currents
in chiral perturbation theory and apply~\cite{Donoghue:1992dd}
\beq
\left<0\right|j^{\mu\,a}_5\left|\pi^b(p)\right> = if_{\pi}p^{\mu}\delta^{ab} \ ,
\eeq
with $f_{\pi} = 93\,\mev$.
This gives the one-pion width
\beq
\Gamma(A\to H\pi^0) = \frac{1}{128\pi}\frac{g^4}{m_W^2}f_{\pi}^2\frac{p}{m_A^2}
(m_AE_{\pi}+E_HE_{\pi}+p^2) 
~\simeq~ \frac{1}{32\pi}\frac{g^2}{m_W^2}f_{\pi}^2\Delta^3 \ ,
\eeq
where $E_i =\sqrt{m_i^2+p^2}$ and
\beq
p^2 = \frac{1}{4m_A^2}(m_A^4+m_H^4+m_{\pi}^4
-2m_A^2m_H^2-2m_A^2m_{\pi}^2-2m_H^2m_\pi^2) \ .
\eeq
The same expression applies to $\Gamma(H^+\to H\pi^+)$
after replacing $\Delta^0\to \Delta^{\pm}$ and $m_A \to m_{H^\pm}$
and adding a factor of $|V_{\rm CKM}^{ud}|^2$.

\begin{figure}[ttt]
 \begin{center}
         \includegraphics[width = 0.55\textwidth]{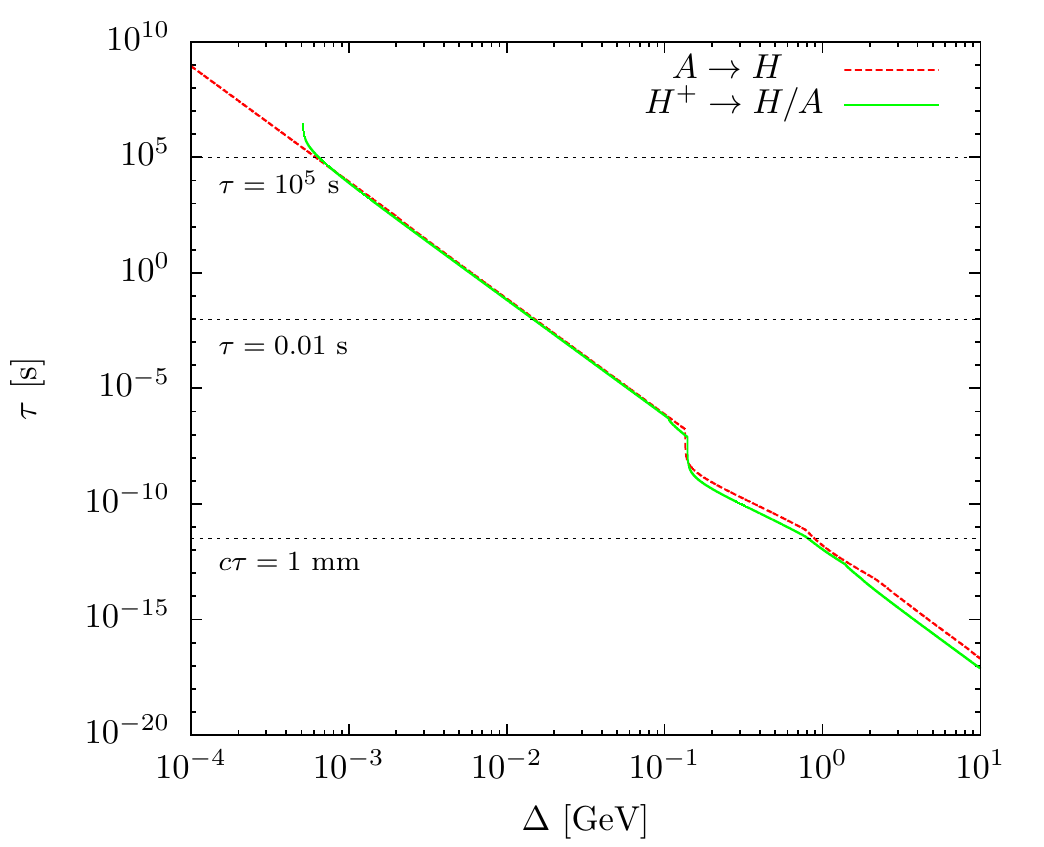}
 \end{center}
\vspace{-0.5cm}
 \caption{Decay lifetimes of the neutral $A$ and charged $H^\pm$
states as a function of the mass splitting $\Delta$ with the lightest
$H$ state.}
 \label{fig:tau}
 \end{figure}

The total lifetimes of $A$ and $H^+$ as a function of the mass splittings
are shown in Fig.~\ref{fig:tau}.  
The mass of $H$ is set to $m_H=70\,\gev$ in generating this figure, 
but the lifetimes are not sensitive to this value
provided $m_H \gg \Delta$.  For $H^+$ decays, we assume 
$\Delta^\pm \gg \Delta^0$ and include decays to both $H$ and $A$ final states.
We also consider only $\Delta^\pm > m_e$ since otherwise $H^+$ is
stable (and in any case, the results of the previous section suggest 
that splittings this small are highly unlikely).  For neutral decays, 
we take $\Delta^\pm > \Delta^0$ and allow for much smaller neutral 
splittings.  The neutral decay width below $\Delta^0 < 2m_e$ is due to neutrino 
final states.\footnote{Note as well that the radiative decay $A\to H+\gamma$ 
is forbidden by angular momentum conservation given the point interaction of
Eq.~\eqref{eq:leff}, in contrast to the decay $\chi_2^0\to \chi_1^0\gamma$
that can occur for neutralinos in supersymmetric theories~\cite{Haber:1988px,Bramante:2014dza}.}  
For both the neutral and charged states, we see from Fig.~\ref{fig:tau}
that the decays are prompt on collider time scales, $c\tau < 1\,\text{mm}$,
for mass splittings above a few hundred MeV.  Some potential implications
of a long-lived neutral state for cosmology will be discussed below.

\section{Basic Bounds\label{sec:basic}}

  The parameters of the IDM are constrained by a number 
of direct and indirect bounds.  By assumption, the model
should produce a locally stable 
electroweak vacuum with only $H_1$ obtaining a VEV.  
This implies~\cite{Barbieri:2006dq,Cao:2007rm}
\beq
\lambda_{1,2} > 0 \ ,~~~~~~~\lambda_3,\,
(\lambda_3+\lambda_4-|\lambda_5|) > -2\sqrt{\lambda_1\lambda_2} \ .
\eeq
The standard electroweak vacuum will also be globally stable if
\beq
\mu_1^2/\sqrt{\lambda_1}< \mu_2^2/\sqrt{\lambda_2} \ .
\eeq
Note that $\mu_1^2 < 0$ is typically needed for EWSB.
It is also standard to demand that the scalar couplings be small enough
to maintain perturbativity to at least a few times the weak scale.
For masses of the charged and neutral scalars that are relatively
close to one another, corresponding to small $\lambda_4$ and $\lambda_5$,
these conditions are met easily for moderate values of $\lambda_3$
and positive values of $\mu_2^2$.

  New electroweakly-charged states can modify precision electroweak
observables.  The leading indirect effect can be described as a shift 
in $\Delta T$ relative to the SM (including a SM Higgs of mass $m_h = 125\,\gev$),
\beq
\Delta T = \frac{1}{32\pi^2\alpha v^2}\left[
F(m_{H^\pm}^2,m_A^2)+F(m_{H^\pm}^2,m_H^2)-F(m_{A}^2,m_H^2)\right] \ ,
\eeq
where $F(x,y) = (x+y)/2 - xy\ln(x/y)/(x-y)$.
This shift in $\Delta T$ can usually be approximated by~\cite{Lundstrom:2008ai}
\beq
\Delta T \simeq \frac{1}{24\pi^2\alpha v^2}(m_{H^\pm}-m_A)(m_{H^\pm}-m_H) \ .
\eeq
This is clearly suppressed in the compressed limit we are considering,
and does not produce a relevant constraint given the currently allowed
range $T = 0.01\pm 0.12$~\cite{Agashe:2014kda}.

  A more direct electroweak requirement is that the SM $W^{\pm}$ and $Z^0$ 
vector bosons should not be able to decay significantly to the exotic scalars.  
Barring extreme kinematic suppression, this implies
\beq
m_H+m_A > m_Z \ ,~~~~~~~m_A+m_{H^\pm},\;m_H+m_{H^\pm} > m_W \ .
\eeq
The limit for $Z$ decays comes from direct searches 
for $Z\to f\bar{f}+\nu\bar{\nu}$, and bounds the process
$Z\to HA$ with $A \to Z^*H$ with $Z^*\to f\bar{f}$~\cite{Cao:2007rm}.  
A similar but slightly weaker bound can be obtained from the invisible decay
width of the $Z$, and is applicable when both $A$ and $H$ are
long-lived, stable, or have very soft decay products~\cite{Cao:2007rm}.  
Limits from $W$ decays to $A H^\pm$ or $H H^\pm$ can presumably be
obtained as well, but will be weaker than the $Z$-decay bound for
$\Delta^{\pm} > \Delta^0$.

  Decays of the SM-like Higgs to light exotic scalars can easily
overwhelm the narrow Higgs width to SM modes.  The partial Higgs decay rate
to a light scalar is~\cite{Cao:2007rm}
\beq
\Gamma(h\to SS) = \xi_S\lrf{m_S^2-\mu_2^2}{m_h}\sqrt{1-4m_S^2/m_h^2} \ ,
\eeq
where $S = H,A,H^\pm$ and $\xi_S = 1\,(2)$ for $S=H,A\;(H^\pm)$.
A recent analysis of LHC Higgs (rate) data applied to the IDM finds
\beq
|m_H^2-\mu_2^2|/2v^2 = |\lambda_3+\lambda_4+\lambda_5|/2 ~\lesssim~ 0.012~(0.007)\ ,
\label{eq:hbound}
\eeq
assuming that only $h\to HH$ is open and 
that $m_H = 60\,\gev~(10\,\gev)$~\cite{Belanger:2013xza}.
When more than one channel is accessible, we expect a corresponding bound to apply
to the orthogonal sum of the effective couplings.  Note that the RG evolution
of $\lambda_3$ is inhomogeneous and sourced by gauge interactions (much like
$\lambda_4$), and one would typically expect values of $\lambda_3$
that are too large to satisfy the condition of Eq.~\eqref{eq:hbound}.

  Even with $m_H > m_h/2$, the IDM can alter the decays of the SM
Higgs boson through loop effects.  In particular, the charged scalar 
modifies the width for $h\to \gamma\gamma$ 
at one-loop with the contribution to the amplitude depending on $\lambda_3$
and $m_{H^{\pm}}$~\cite{Cao:2007rm}.  This only provides a moderate bound on 
$|\lambda_3| \lesssim 1$ 
for $m_{H^{+}} > m_Z/2$~\cite{Belanger:2013xza,Arhrib:2012ia,Swiezewska:2012eh}.

\section{Cosmological Constraints\label{sec:cosmo}}

  The IDM contains a promising dark matter~(DM) candidate provided
the lightest new scalar state is neutral.    
A compressed mass spectrum in the IDM has important implications
for the relic density and detection prospects of this DM component.
Very small mass splittings can also lead to long-lived metastable
states whose late decays may be cosmologically important.
We examine these topics in this section.

\subsection{Relic Abundances\label{sec:abun}}

  Thermal dark matter production in the IDM has been studied extensively~\cite{
Barbieri:2006dq,LopezHonorez:2006gr,Cao:2007rm,Dolle:2009ft,
Hambye:2009pw,Goudelis:2013uca,
Dolle:2009fn,Honorez:2010re,LopezHonorez:2010tb,Krawczyk:2013jta,
Arhrib:2013ela, Ilnicka:2015jba}.  
We extend these works by elucidating the key dynamics
in the compressed regime.  An important implication of the compressed
spectrum is that coannihilation plays a central role in determining the 
relic density of the lightest state, which we assume to be $H$.
Again, for our analysis we assume a small mass splitting between the neutral
DM states with $\Delta^0 \sim 100\; \kev$ (motivated by 
direct detection constraints in Sec.~\ref{sec:dd}), but we consider 
larger mass splittings between the charged and neutral states, 
$\Delta^{\pm} \geq 1\,\gev$. 
As shown in Sec.~\ref{sec:split}, the coupling $\lambda_5$ and 
therefore $\Delta^0$ can be taken to be arbitrarily small 
in a technically natural way. 
The specific value of $\Delta^0$ does not matter much for the relic density
provided it is much smaller than the freeze-out temperatures we obtain
\footnote{For $m_H>1\;\GeV$ typical freeze-out temperatures are 
$\Tfo\sim m_H/20 \gg \Delta^0$.},
since in this case coannihilation between the $H^0$ and $A^0$ states
will not receive any significant Boltzmann suppression.  In contrast,
the charged-neutral splitting $\Delta^{\pm}$ is sourced by gauge couplings
in the course of RG running and can be significantly larger.
Thus, we examine the effects of larger $\Delta^{\pm}$ on the DM
relic density.

\begin{figure}[ttt]
\centering
\includegraphics[width=0.47\textwidth]{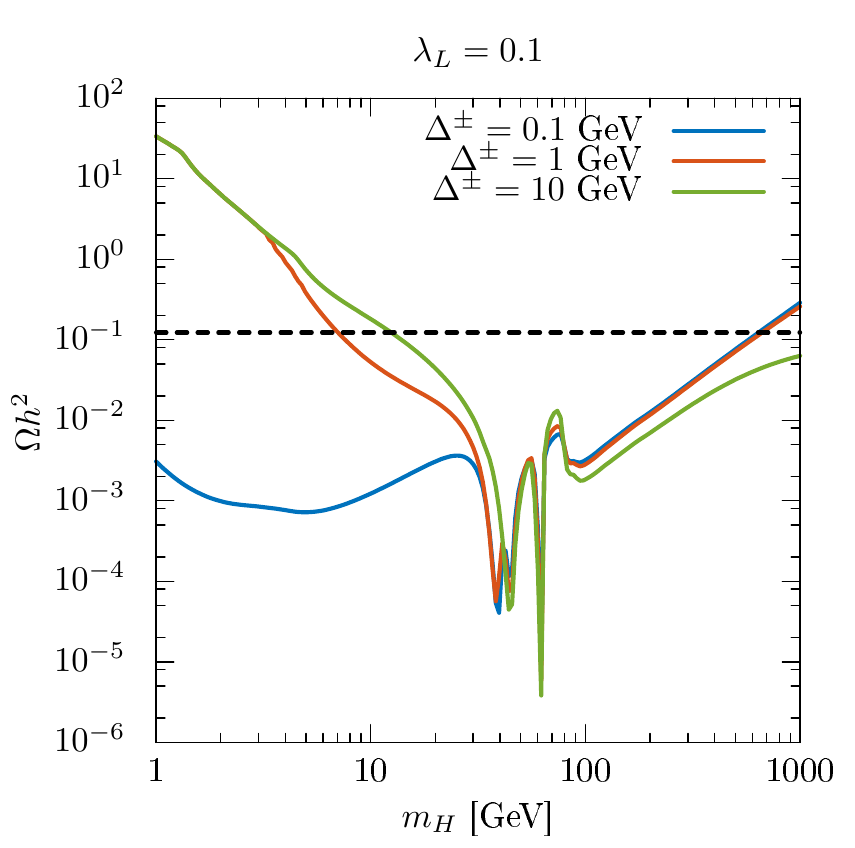}
\includegraphics[width=0.47\textwidth]{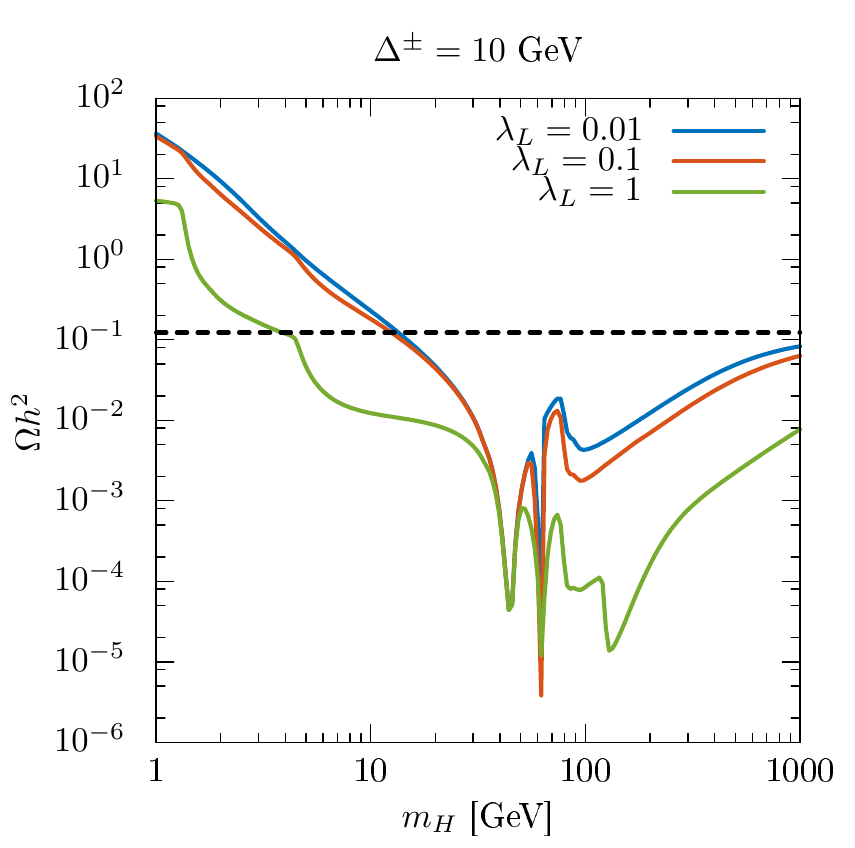}
\caption{Abundance of IDM dark matter as a function of mass 
in the compressed regime.  The left panel shows the abundance for
fixed $\lambda_L=0.1$ and charged-neutral mass splittings of 
$\Delta^\pm=0.1,\,1.0,\,10\,\gev$.  The right panel shows the relic abundance
for $\lambda_L=0.01,0.1,1.0$ and fixed splitting $\Delta^\pm =10\;\GeV$. 
The horizontal dashed line in both panels is the 1$\sigma$ upper limit 
on the total dark matter density as determined by the Planck 
collaboration~\cite{Ade:2015xua}.
\label{fig:abun}}
\end{figure}

We compute the relic abundance of $H$ as a function of its mass for 
various choices of the mass splittings $\Delta^0$ and $\Delta^\pm$ and the
Higgs coupling parameter
$\lambda_L = (\lambda_3+\lambda_4+\lambda_5)/2$.
The calculation is performed using \micromegas~4.1.8\cite{Belanger:2013oya}, 
which includes the effects of coannihilation as well as two-, three-, and four-body 
annihilation channels. Our main results are summarized 
in Fig.~\ref{fig:abun}.  The left panel of this figure shows 
the impact of the charged-neutral splitting 
$\Delta^\pm$ on the abundance for fixed $\lambda_L$, while the right 
panel shows the effect of varying $\lambda_L =  0.01,\;0.1,\;1$
with fixed $\Delta^{\pm} = 10\,\gev$. 
Note that for $m_H \leq m_h/2$, these sizable $\lambda_L$ will typically
induce an unacceptable $h\rightarrow\text{invisible}$ decay rate
as discussed in Sec.~\ref{sec:basic}, Eq.~\eqref{eq:hbound}. 
Also shown in both panels, by the horizontal dashed line, is the
$1\sigma$ upper limit on the dark matter density 
$\Omega_{\rm cdm} h^2\lesssim 0.1227$ 
as determined by the Planck collaboration~\cite{Ade:2015xua}.

  The relic abundances in Fig.~\ref{fig:abun} can be understood
in terms of coannihilation, resonances, and Higgs couplings.
Coannihilation usually relies on gauge couplings, so when it is important 
the abundance curves for different values of $\lambda_L$ collapse. 
This can be seen near the $Z$ resonance at $m_H = m_Z/2$
where $A$-$H$ coannihilation dominates.  A Higgs funnel at $m_H = m_h/2$ 
that depends on $\lambda_L$ is also visible in Fig.~\ref{fig:abun}. 
The relic abundance below $m_H \lesssim 40\,\gev$ is very sensitive
to the mass splitting $\Delta^{\pm}$, as can be seen in the left panel
of Fig.~\ref{fig:abun}.  Here, coannihilation with $H^\pm$
plays a leading role for small mass differences because
the charged state has the direct two-body annihilation channel
$H^+H^-\to \gamma\gamma$ open, while the neutral states must go
through three- and four-body channels in this region.

At higher masses, the dominant annihilation process is into pairs of weak
vector bosons~\cite{Hambye:2009pw}.  
Annihilation into off-shell vector bosons can also dominate
below threshold~\cite{Honorez:2010re}, 
(and is included in the version of 
\micromegas~4.1.8~\cite{Belanger:2013oya} we use).
The relative contributions of transverse and longitudinal final states
depend on the scalar couplings $\lambda_{i>2}$, 
with annihilation to longitudinal modes enhanced for larger values 
of these couplings~\cite{Hambye:2009pw} as expected from 
the Goldstone equivalence theorem~\cite{Lee:1977eg,Chanowitz:1985hj}.  
In the compressed regime, $|\lambda_4|$ and $|\lambda_5|$ are both assumed to be 
small, leaving $\lambda_3\simeq 2\lambda_L$ as the only potentially large one.
The enhancement of the annihilation rate with larger $\lambda_L$
(and to a lesser extent with larger $\Delta^\pm$ corresponding to
increased $|\lambda_4|$) is clear in Fig.~\ref{fig:abun}.
When all the scalar couplings are small, the observed relic density
is obtained from $m_H \simeq 535\,\gev$~\cite{Hambye:2009pw}.

\subsection{Dark Matter Detection\label{sec:dd}}

  The analysis above shows that for $m_Z/2 \lesssim m_H \lesssim 500\,\gev$,
the relic $H$ abundance can only make up a fraction of the total DM density
when the scalar spectrum is compressed.
Despite this, direct detection limits place strong additional constraints
on the IDM relic component.  Indirect detection does not appear to provide
any further constraint.

  Direct detection rules out the IDM for neutral mass splittings
with $\Delta^0 \lesssim 100\,\kev$ for $m_H > m_Z/2$~\cite{Barbieri:2006dq}.  
In this highly degenerate limit, 
there is unsuppressed inelastic spin-independent scattering with nuclei, 
$HN\to AN$~\cite{TuckerSmith:2001hy,Arina:2009um}, 
via a vector interaction through the $Z$ 
with a very high cross section.  
This process turns off for neutral splittings above 
$\Delta^0 \gtrsim 200\,\kev$, and we will
assume values at least this large for the remainder of the analysis.

In the presence of such a splitting between $H$ and $A$, 
the dominant interaction between IDM relics in the halo and nuclei is
spin-independent~(SI) elastic scattering through SM Higgs exchange. 
The resulting DM-nucleon cross-section is~\cite{Majumdar:2006nt,Kanemura:2010sh}
\beq
\sigma_n = \frac{\lambda_L^2}{\pi}\lrf{m_n}{m_H+m_n}^2\frac{f_n^2}{m_h^4} 
\ ,
\label{eq:ddxsec}
\eeq
where
\beq
f_n = \sum_{q=u,d,s} m_q \langle n | \bar q q | n \rangle - \frac{\alpha_s}{4\pi} \langle n | G_{\mu\nu}^a G^{a\mu\nu} | n \rangle \ .
\eeq
These matrix elements can evaluated using the results of Ref.~\cite{Hill:2014yxa}.

\begin{figure}
\centering
\includegraphics[width=0.47\textwidth]{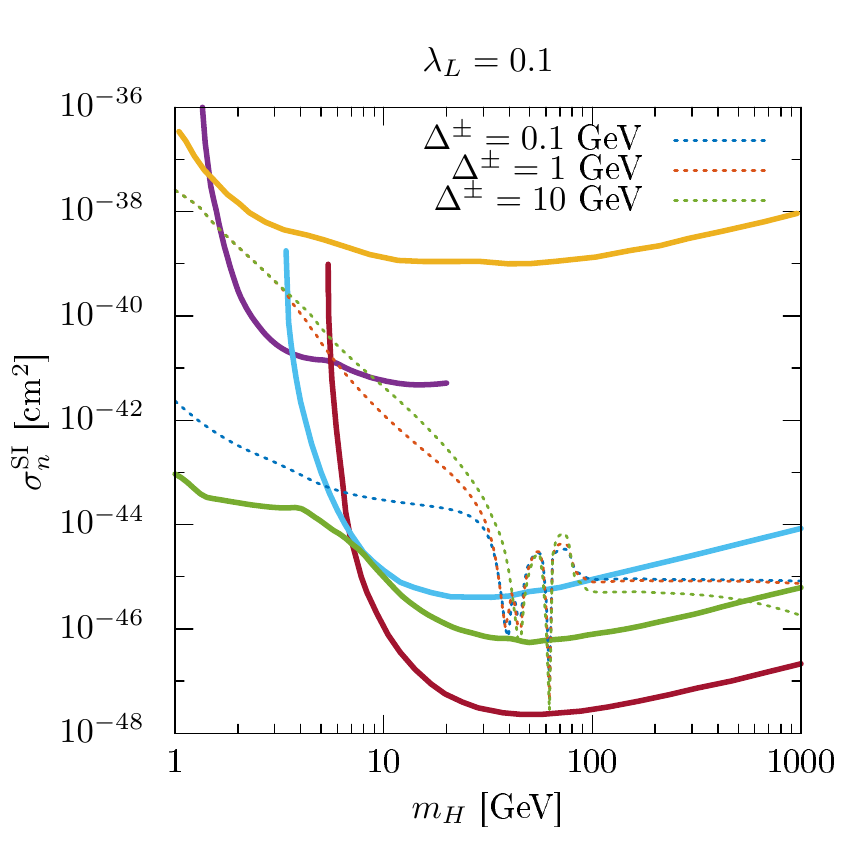}
\includegraphics[width=0.47\textwidth]{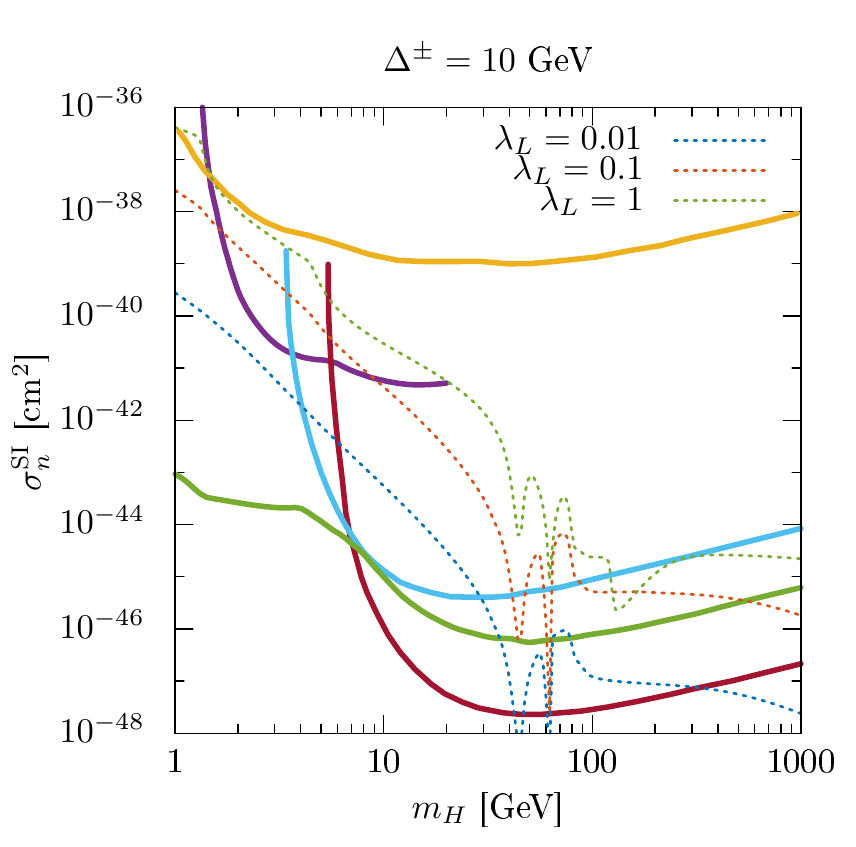}
\caption{Limits on the effective DM-nucleon cross section as a function 
of mass from direct detection experiments and predictions for the IDM
relic scaled by the expected thermal relic density.
The solid lines correspond to the bounds from 
LUX~\cite{Akerib:2015rjg}~(light blue),
CDMSLite~\cite{Agnese:2015nto}~(purple), 
CRESST-Si~\cite{Angloher:2002in}~(yellow)
and projected sensitivities of LZ~(red)~\cite{Akerib:2015cja} 
and SuperCDMS SNOLAB~(green)~\cite{Cushman:2013zza}.
The dashed lines show the nucleon-scattering cross
sections of $H$ rescaled by $\Omega/\Omega_{\rm cdm}$, 
with the colour coding identical to Fig.~\ref{fig:abun}. 
\label{fig:dd}}
\end{figure}

The resulting bounds on the IDM relic $H$ from spin-independent 
nucleon scattering
are shown in Fig.~\ref{fig:dd} as functions of the mass $m_H$.   
In both panels, the solid lines indicate the current limits from direct
detection experiments, with the best current bounds for $m_H\gtrsim 5\;\GeV$ 
from the LUX experiment~\cite{Akerib:2015rjg}~(light blue),
along with  
CDMSLite~\cite{Agnese:2015nto}~(purple) 
and CRESST-Si~\cite{Angloher:2002in}~(yellow) dominating at lower masses.
We also show the projected sensitivities of LZ~(red)~\cite{Akerib:2015cja} 
and SuperCDMS SNOLAB~(green)~\cite{Cushman:2013zza}.
The dashed lines in Fig.~\ref{fig:dd} show the nucleon-scattering cross
sections of $H$ rescaled by $\Omega/\Omega_{\rm cdm}$, 
with the colour coding identical to Fig.~\ref{fig:abun}. 
Comparing these contours, we see that a light IDM relic component
with $m_Z/2 < m_H < 100\,\gev$ can be consistent with current bounds 
from direct detection.  However, these bounds also suggest that
$|\lambda_L|$ should be much smaller than unity in this mass region
unless the relic density of $H$ is particularly small.  Constraints
from direct detection are easier to satisfy for larger $H$ masses,
including $m_H\gtrsim 500\,\gev$ where the IDM relic can potentially make up
the full DM relic density.

In contrast to direct detection, indirect detection signals of the IDM relic 
for $m_H \lesssim 500\,\gev$ are typically well below current 
sensitivities~\cite{Arhrib:2013ela,Gustafsson:2007pc,Modak:2015uda}.  
This is mainly due to the sub-critical density of 
$H$, but also from the much greater effectiveness of coannihilation 
during freeze out compared to today.  Capture and annihilation of the $H$
state in the Sun or Earth is more promising~\cite{Agrawal:2008xz,Andreas:2009hj},
but limits from direct detection are typically 
much stronger~\cite{Andreas:2009hj}.

\subsection{Late-Time Decays}

  The heavier IDM scalars can
become long-lived on cosmological timescales for small mass splittings.
From Fig.~\ref{fig:tau}, we see that the decay lifetimes exceed 
$\tau \gtrsim 0.01\,\text{s}$ for splittings below $\Delta \lesssim 20\,\mev$.
Such decays can disrupt primordial nucleosynthesis~(BBN)
and inject additional energy into the cosmic microwave background~(CMB).

  Mass splittings below $\Delta \lesssim 20\,\mev$ are only expected
for the neutral states, so we will focus on $A\to HZ^*$ modes.
The dominant final states in this case are $e^+e^-$ and $\nu\bar{\nu}$,
with no hadronic component present.  
Comparing to the direct bounds
on late-time electromagnetic energy injection 
from Ref.~\cite{Jedamzik:2004er,Kawasaki:2004qu},
which only become significant for $\tau > 10^5\,\text{s}$,
we do not find a limit from BBN. Additionally, the energy
released per decay is usually too small to destroy light nuclei.
Very late decays with $\Delta^0 < 1\,\mev$ will produce mostly
neutrinos, but can also yield photons through loops.  
These photons can disrupt the black body spectrum of the CMB~\cite{Hu:1993gc}.
However, given the expected primordial yield of $A$ and photonic
branching fraction, the analysis of Ref.~\cite{Kanzaki:2007pd} indicates 
that no CMB limit is obtained for $\Delta^0 > 100\,\kev$, as required
by direct detection.

\section{LEP Limits\label{sec:lep}}

  Data from LEP can probe the IDM beyond the basic bounds outlined 
in Section~\ref{sec:basic}.  The dominant production modes for the new
scalars at LEP are $e^+e^-\to H^+H^-$ via a $\gamma$ or $Z$, and 
$e^+e^-\to HA$ via a $Z$.  Once created, the heavier states will
decay down to the lightest neutral state (assumed to be $H$ here)
through $H^{\pm}\to HW^{\pm*}$ and $A\to HZ^{*}$, with $H$
escaping the detector as missing energy.  These features are very
similar to the signals of supersymmetry in $\chi_1^+\chi_1^-$ and
$\chi_2^0\chi_1^0$ production, and thus LEP searches for electroweak
superpartners are natural to apply to the IDM.  

  A detailed analysis of this kind was undertaken 
in Ref.~\cite{Lundstrom:2008ai}, where LEP searches for 
$\chi_1^0\chi_2^0$ in the DELPHI experiment~\cite{Delphi1} were used
to derive an exclusion on $HA$ production.\footnote{See also 
Refs.~\cite{Barbieri:2006dq,Cao:2007rm}.}  
The limits found in this work extend up to $m_A \simeq 100\,\gev$
for larger neutral mass splittings, but there is no improvement over 
the $Z$ decay bound ($m_A+m_H> m_Z$) for $\Delta^0 < 8\,\gev$.
For the charged states, a reinterpretation of the OPAL results of
Refs.~\cite{Abbiendi:2003ji,Abbiendi:2003sc} was made 
in Ref.~\cite{Pierce:2007ut} to derive a limit
of $m_{H^{\pm}} \gtrsim 70\!-\!90\,\gev$.  However, this analysis
was approximate, and is only applicable to mass differences 
$\Delta^\pm \gtrsim 5\,\gev$.  

  In this section we refine the estimate of the limits from LEP
on $H^+H^-$ production by reinterpreting searches for $\chi_1^+\chi_1^-$
production.  We also apply LEP searches using monophotons
to test highly-compressed IDM spectra.  Throughout our analysis,
we assume $\Delta^\pm \gg \Delta^0$ and $\Delta^0 \lesssim 0.5\,\gev$.
This implies that the $H$ and $A$ states are not constrained by
the LEP reanalysis of Ref.~\cite{Lundstrom:2008ai}, that the products
of $A\to HZ^*$ decays are largely invisible to the LEP detectors,
and that the $H^+\to AW^*$ and $H^+\to HW^*$ modes are effectively
indistinguishable.

\subsection{General Searches for $H^{+}H^-$}

Experiments at LEP looked for supersymmetric chargino
$\chi_1^+\chi_1^-$ production in final states with leptons or jets 
and missing energy.  
The OPAL search of Ref.~\cite{Abbiendi:2003ji}, using $680\,\text{pb}^{-1}$
of data at center-of-mass energies of 183-209~GeV,
focused on dileptons and missing energy, with each chargino assumed 
to decay through an off-shell $W$ via $\chi_1^+\to \chi_1^0W^*$.  
Exactly the same decay topology occurs for $H^+H^-$ production in the IDM,
with $H^+\to HW^*$ and $H^+\to AW^*$, and thus the OPAL result can be used 
to constrain the IDM as well.  
Similar chargino searches were performed by the other LEP 
collaborations~\cite{Abbiendi:2003sc,Abdallah:2003xe}, 
and were combined in Ref.~\cite{lepchg1}, 
but only Ref.~\cite{Abbiendi:2003ji} stated their results 
with enough resolution to be useful for the present analysis.

  The OPAL search of Ref.~\cite{Abbiendi:2003ji} concentrated on 
final states with dileptons and missing energy.  Their exclusions are
given in terms of limits on the inclusive $\chi_1^+\chi_1^-$ production 
cross section at $\sqrt{s}=208\,\gev$ times the square of the branching 
fraction $BR(\chi_1^+\to\chi_1^0{\nu}_{\ell}\ell^+)$, with data taken
at different collision energies re-weighted with a factor
of $\beta/s$ (where $\beta = \sqrt{1-4m^2_{H^\pm}/s}$).  
To translate these limits to the IDM, 
we calculate the total $H^+H^-$ production cross section
at $\sqrt{s} = 208\,\gev$ and the leptonic branching fraction
$BR(H^+\to H{\nu}_{\ell}\ell^+)$ in \texttt{Madgraph5}~\cite{Alwall:2014hca}.  
For $\Delta^{\pm}\gg \Delta^0$, which we assume here, 
the same branching ratio is expected for $H^+\to AW^*$.  
To account for the different threshold behaviour of scalar production
($\beta^3/s$) relative to fermion production ($\beta/s$),
we apply an additional re-weighting factor to account for the 
multiple collision energies used in the OPAL analysis~\cite{Abbiendi:2003ji}.
We then apply the limits quoted in Ref.~\cite{Abbiendi:2003ji}
to the corresponding combination in the IDM to derive exclusions
on the IDM parameter space.

\begin{figure}[ttt]
 \begin{center}
         \includegraphics[width = 0.55\textwidth]{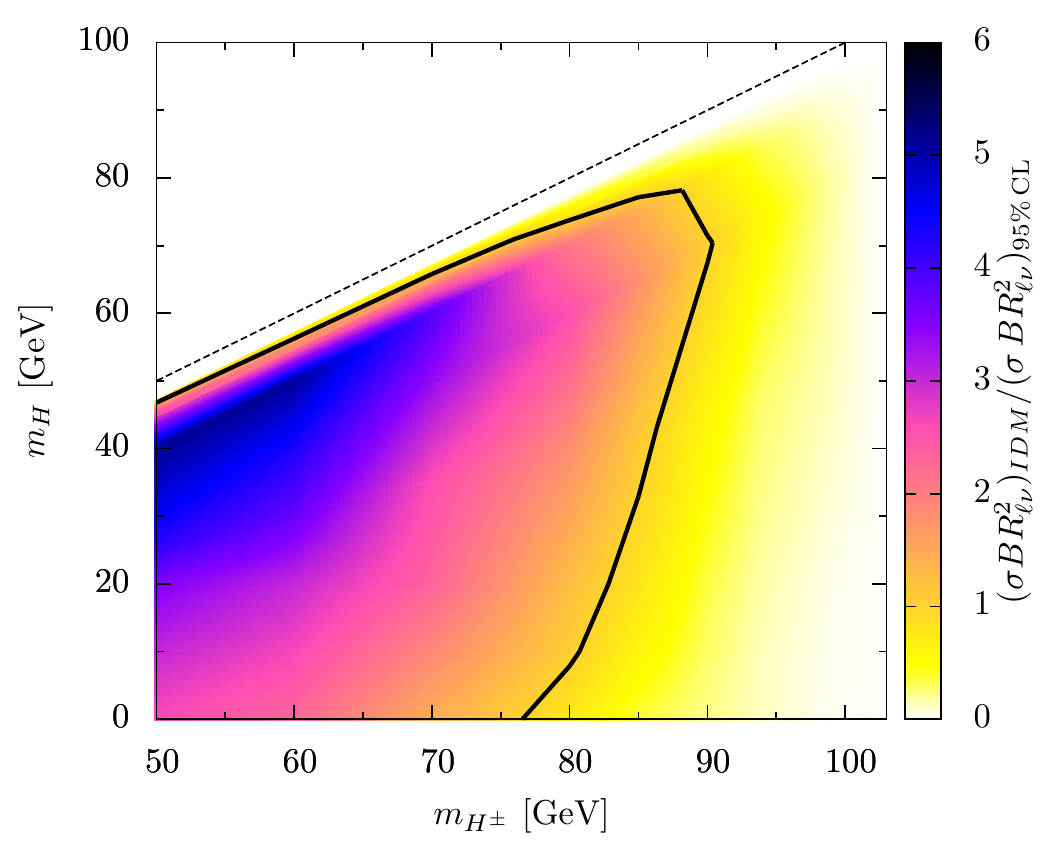}
 \end{center}
\vspace{-0.5cm}
 \caption{Excluded region ($95\%\,\text{CL}$) in the $m_{H^\pm}\!-\!m_H$
plane derived from the OPAL search for leptons and missing energy
of Ref.~\cite{Abbiendi:2003ji}. The colour map shows the magnitude of 
$(\sigma BR^2_{\ell\nu})_{IDM}/(\sigma\,BR^2_{\ell\nu})_{95\%\,\text{CL}}$. }
 \label{fig:chg-nd}
 \end{figure}

  Figure~\ref{fig:chg-nd} shows the $95\%\,\text{CL}$ exclusion derived from 
our reanalysis in the $m_{H^\pm}\!-\!m_H$ plane by the solid black contour,
with the colour map showing the ratio of the predicted signal rate 
to the excluded value, 
$(\sigma BR^2_{\ell\nu})_{IDM}/(\sigma\,BR^2_{\ell\nu})_{95\%\,\text{CL}}$.
The diagonal dashed line indicates $m_{H^\pm} = m_H$ degeneracy. 
The exclusion extends all the way up to $m_{H^+} \simeq 90\,\gev$ 
for moderate mass splittings,
in agreement with the analysis of Ref.~\cite{Pierce:2007ut}, but falls off 
sharply in the degenerate limit of $\Delta^{\pm} \lesssim 5\,\gev$
due to the reduced detection efficiencies for soft leptons.
The IDM exclusion is also somewhat weaker than for charginos due
to the lower relative production cross section.

It should be noted that our analysis assumes implicitly
that the detection efficiencies are the same for the $H^+H^-$ signal
as for $\chi_1^+\chi_1^-$ when the charged and neutral masses match up.
We expect this to be a good approximation based on the analysis
of Ref.~\cite{Lundstrom:2008ai}, which found that the 
efficiencies for a similar LEP search by the DELPHI collaboration 
for $\chi_2^0\chi_1^0$ production followed by $\chi_2^0\to \chi_1^0Z^*$ 
with $Z^*\to f\bar{f}$ were equal to those
of the corresponding IDM process $A\to H Z^*$ with $Z^*\to f\bar{f}$ 
to within $10\!-\!20\%$ throughout the relevant parameter space.

A full recasting~\cite{Cranmer:2010hk} of the LEP analyses using the 
combined data from all four LEP experiments and including hadronic searches
could provide a slightly stronger bound on the IDM mass spectrum.  
At the same time, the limits derived from such an analysis are 
unlikely to alter our general conclusions.  
Namely, we expect the compressed region of $\Delta^\pm \lesssim 5\,\gev$
to remain unconstrained due to the difficulties in identifying 
objects with such small visible energies.  
To address this region, we turn next to LEP analyses 
designed to find charginos that are nearly degenerate with 
a lightest neutralino superpartner and apply them to the IDM.

\subsection{Compressed Searches for $H^+H^-$ with a Photon}\label{sec:compressed}

\begin{figure}[ttt]
 \begin{center}
         \includegraphics[width = 0.55\textwidth]{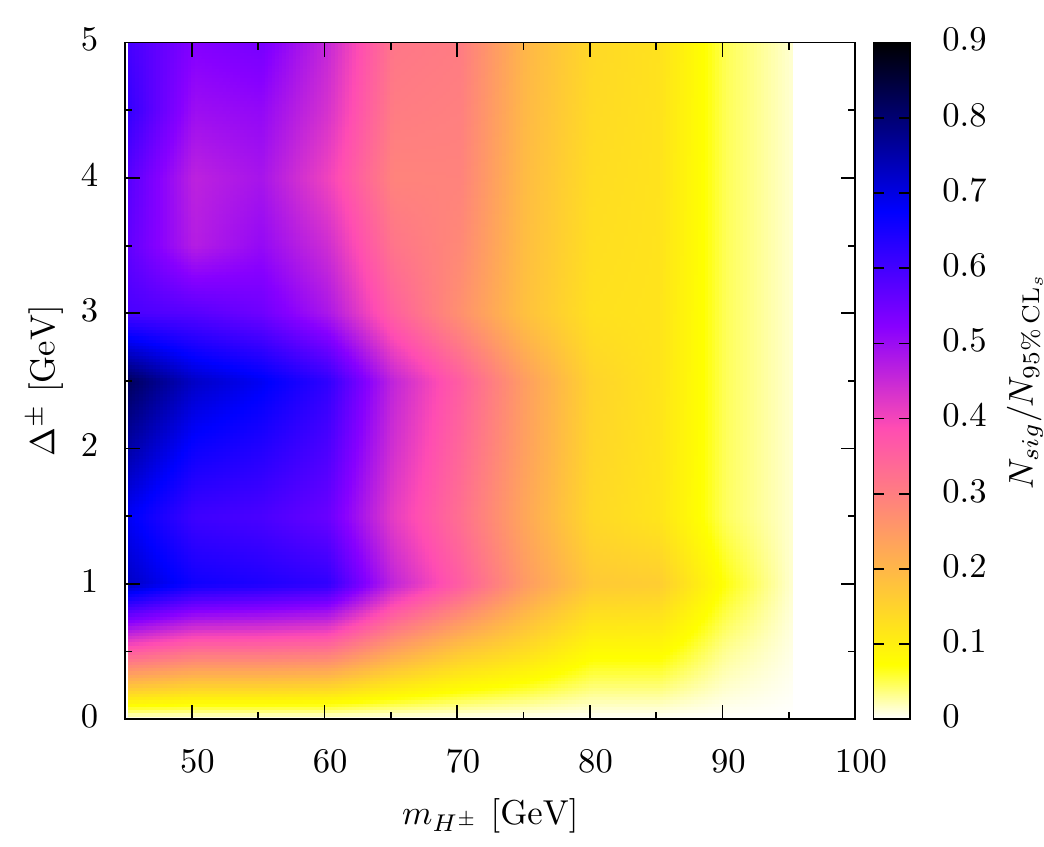}
 \end{center}
\vspace{-0.5cm}
\caption{Magnitude of $\sigma(e^{+}e^{-}\to H^{+}H^{-}\gamma)/\sigma^{95\%~C.L}_{\text{OPAL}}$~\cite{Abbiendi:2002vz} in the $m_{H^{+}}-m_{H}$ plane.}
 \label{fig:chg-dg}
 \end{figure}

Mass splittings below $\Delta^\pm \lesssim 5\,\gev$ result in $H^\pm$ decays
that leave very little hadronic or leptonic activity in the detector 
and that are difficult to distinguish from background.  
Some sensitivity can be regained by studying events with soft charged 
tracks together with an additional photon of moderate transverse momentum $p_T$ 
from initial- or final-state radiation.  
Searches of this type have been carried out 
in Refs.~\cite{Abdallah:2003xe,Acciarri:2000wy,Abreu:2000as,Heister:2002mn,Abbiendi:2002vz},
and are summarized in Ref.~\cite{lepchg2}.  In this section, we apply the OPAL
analysis of Ref.~\cite{Abbiendi:2002vz} to the compressed IDM
with charged mass splittings $\Delta^\pm \lesssim 6\,\gev$.

The OPAL search of Ref.~\cite{Abbiendi:2002vz} was designed to probe
nearly mass-degenerate charginos and neutralinos, and used $570\,\text{pb}^{-1}$
of data with center-of-mass energies between 189--208~GeV.  
Potential signal events were selected if they contained a high-energy photon 
with $E_T > 5\,\gev$ and $|\cos\theta^{\gamma}|< 0.976$ together with
two to ten high quality tracks, with further cleaning and final selection
cuts applied subsequently.  To estimate the IDM signal in this search,
we use \texttt{Madgraph5}~\cite{Alwall:2014hca} to generate $H^+H^-$ events 
in association with a photon having $E^{\gamma}>0.025\sqrt{s}$ 
and $|\cos\theta^{\gamma}|>0.985$ from $e^+e^-$ collisions at $\sqrt{s}=208$ GeV.
We then apply the efficiencies quoted in Ref.~\cite{Abbiendi:2002vz}
for Higgsino-like charginos with the same photon requirements
(Fig.\,4c in Ref.~\cite{Abbiendi:2002vz}).\footnote{
Note that by applying the efficiencies to the subset of events with an
additional hard photon, potential differences in the photon radiation
spectrum between Higgsinos and the IDM scalars are taken into account.}
To extrapolate these efficiencies to the entire $m_{H^\pm}\!-\!\Delta^{\pm}$ 
plane, we assume a sharp linear rise from 
$\Delta^{\pm}=0$ to $\Delta^{\pm}=1\,\gev$
and a flat efficiency for $\Delta^\pm\geq 1\,\gev$ 
(Fig.\,4d of Ref.~\cite{Abbiendi:2002vz}).  As in the general
chargino search considered above, our analysis assumes that the efficiencies
for this subset of $H^+H^-$ events will be very similar to the corresponding
subset of chargino events.  We expect this to be a reasonable approximation,
but a full recasting would provide a more definitive answer.

  With this estimate of the IDM signal, we attempt to derive limits 
on the theory in the compressed regime by comparing our prediction 
to the number of observed and expected background events in the OPAL search 
(Fig.\,3a,b of Ref.~\cite{Abbiendi:2002vz}).  Applying a $\text{CL}_{s}$ 
statistic, no limit is found at $95\%\,\text{CL}_s$.  
In Fig.~\ref{fig:chg-dg} we show the
ratio of the number of estimated $H^+H^-$ signal events to 
the upper limit allowed by the data, $N_{sig}/N_{95\%\,\text{CL}_s}$,
within the $m_{H^\pm}\!-\!\Delta^{\pm}$ plane.  
While no exclusion is found for $m_{H^\pm} > m_Z/2$, the signal 
does approach the exclusion limit for smaller $m_{H^\pm}$ masses.
It is possible that an exclusion could be derived from a combination 
of data from all four LEP experiments.  The combined limits given in
Ref.~\cite{lepchg2} do not provide enough resolution to allow us to do so.

\subsection{Other LEP Searches}

  A number of other LEP searches can be applied to the compressed IDM
to probe both $H^+H^-$ and $HA$ production.  Very long-lived $H^\pm$
states will leave charged tracks in the LEP detectors.  Searches
for such tracks have been performed 
in Refs.~\cite{Ackerstaff:1998si,Abreu:2000tn,Abbiendi:2003yd}, 
and typically require a track length of at least several $\text{cm}$.
The bounds derived from these searches are quite strict, approaching
the kinematic production limit; the analysis of Ref.~\cite{Abbiendi:2003yd} 
excludes $H^\pm$ masses up to $m_{H^\pm} \lesssim 95\,\gev$. However, lifetimes long
enough to produce charged track lengths longer than a few centimetres 
only occur in the IDM for charged mass splittings below a few hundred MeV 
(Fig.~\ref{fig:tau}), which is well below the typical expected splitting  
from quantum corrections (Fig.~\ref{fig:delm}).  

Various LEP searches can also be applied to $HA$ processes
as in Ref.~\cite{Lundstrom:2008ai}.  In the moderately compressed regime,
with neutral mass splittings in the range $\Delta^0 \in [0.5,\,5]\,\gev$,
the OPAL monophoton plus charged tracks analysis of Ref.~\cite{Abbiendi:2002vz} 
applied to $H^+H^-$ production above will likely have a similar sensitivity
to $HA$ processes.  Smaller mass splittings can be tested using
pure monophoton searches (with a veto on charged tracks), and have been
considered in Refs.~\cite{Abdallah:2003np,Abdallah:2008aa}
and applied to simplified models of dark matter
in Ref.~\cite{Fox:2011fx}.  We find that the cross sections excluded by these
analyses are typically an order of magnitude larger than in the compressed
IDM once the additional photon requirement is imposed.  

  In summary, the results of this section suggest that LEP does not
constrain the compressed IDM for masses above $m_Z/2$ provided the 
charged mass splitting lies below about $\Delta^\pm\lesssim 5\,\gev$
and above a few hundred MeV.  We have also derived a refinement 
of the limits on $H^+H^-$ production for $\Delta^\pm$ larger than 5~GeV.

\section{LHC Limits and Projected Sensitivity \label{sec:lhc}}

As we have seen, LEP searches can constrain the IDM up to masses close
to $90\,\gev$, but they do not provide a bound beyond $m_H > m_Z/2$
when the spectrum is compressed.  In this section, we investigate whether
existing and future searches at the LHC can further probe this region 
of the IDM.
Motivated by the discussion of Sec.~\ref{sec:split}, 
we restrict our attention to the compressed scenario with 
$\Delta^0 = (m_A\!-\!m_H) \sim 100\,\text{keV}\!-\!5\,\gev \ll \Delta^\pm$ and 
$\Delta^{\pm}=(m_{H^\pm}\!-\!m_H) \sim 1\!-\!30\,\gev$ with $m_H>m_Z/2$.
Our general conclusion is that this regime is also very difficult
to test at the LHC, but that mono-jet searches at $\sqrt{s}=14$~TeV 
could be sensitive to lower masses.

The dominant production modes of the $H$, $A$, and $H^{\pm}$ states 
at the LHC are
\beq
q\bar{q} &\to& Z^*~\to~ HA\label{eq:ha}\\
q\bar{q} &\to& \gamma/Z^* ~\to~ H^+H^-\label{eq:hpm}\\ 
q\bar{q} &\to& W^{\pm\,*} ~\to~ H^{\pm}H,\,H^{\pm}A\label{eq:hha}\\
gg &\to& h^{(*)} ~\to~ HH,\,AA,\,H^+H^- \ .\label{eq:hhiggs}
\eeq
The first three processes are electroweak, while the last involves
the additional difficulty of producing a Higgs.  
Production through the Higgs is also proportional to $\lambda_L^2$,
and is suppressed even further when this coupling 
is small~\cite{Craig:2014lda}.  
We will neglect it in the present discussion.\footnote{Small $|\lambda_L|$
is also motivated by the direct detection bounds found in Sec.~\ref{sec:cosmo}.
Note as well that for small $|\lambda_L|$, modifications to the production
and decay rates of the SM Higgs boson $h$ will also be negligible.}
In this limit, the production of inert scalars is very similar to 
that of light degenerate Higgsinos in the MSSM, but with a rate that
is smaller by about an order of magnitude.

\subsection{Exclusive Lepton Searches}

Several previous studies have analyzed leptonic signatures of the IDM 
at the LHC~\cite{Barbieri:2006dq,Cao:2007rm,Dolle:2009ft,Miao:2010rg,Gustafsson:2012aj,Belanger:2015kga}.  Of the various final states considered, 
dilepton searches are typically the most promising.  
However, they are unlikely to provide additional constraints 
on the inert doublet model in the compressed limit.
For example, Ref.~\cite{Belanger:2015kga} used the ATLAS study 
of Ref.~\cite{Aad:2014vma} to constrain the IDM, considering events
with exactly two leptons with $p_T > 35,\,20\,\gev$.  
From these existing results, Ref.~\cite{Belanger:2015kga}
found new limits for $\Delta^0 \gtrsim 65\,\gev$.  However, 
for smaller mass differences, no new limit is found due to 
the lepton $p_T$ requirements of the search.

 More specific IDM searches using dileptons were proposed for the 14\,TeV 
LHC in Refs.~\cite{Cao:2007rm,Dolle:2009ft}. 
Both focus on the production channel of Eq.~\eqref{eq:ha} and make use
of the lepton kinematics expected from their origin in $Z^*\to \ell^+\ell^-$.
Specifically, for small $\Delta^0$ one expects $m_{\ell\ell} \sim \Delta^0$,
small lepton $p_T$, and nearly collinear leptons in most events;
this helps to reduce the SM background.  The parton-level result 
of Ref.~\cite{Cao:2007rm} finds $S/\sqrt{B} \sim 5$ for $100\,\text{fb}^{-1}$ 
of data (with $S/B\sim 1$) for $(m_H,m_A) = (50,\,60)\,\gev$.  
However, Ref.~\cite{Dolle:2009ft} performed a similar analysis and found 
less optimistic results once lepton isolation and detector effects 
were taken into account. It is possible that a lepton-jet type 
search could recover some sensitivity~\cite{Aad:2014yea}, 
but the compressed limit of the IDM is unlikely to be covered 
by these strategies.

Searches with trileptons were considered in Ref.~\cite{Miao:2010rg}
while four-lepton signatures were studied in Ref.~\cite{Gustafsson:2012aj}.  
In both cases, no limits appear to be attainable at the LHC for 
small mass splittings, even at $14\,\tev$ and large integrated luminosity.

\subsection{Mono-Jet Searches}\label{sec:monojet}

Dedicated lepton searches decrease in sensitivity in the degenerate regime 
simply because the resulting leptons are very soft. 
This suggests an alternative detection strategy for the inert scalars 
of the compressed IDM: search for a hard object 
(\textit{e.g.} a jet, Higgs, or gauge boson) produced in association 
with substantial missing energy. For the models of interest, 
the pure mono-jet channel is the most promising. Searches of this type 
typically veto on leptons with $p_T > p_T^{\ell, {\rm min}} \gtrsim 10$ GeV. 
If both $\Delta^{\pm} < p_T^{\ell, {\rm min}}$ and 
$\Delta^{0} < 2 p_T^{\ell, {\rm min}}$, the transverse momentum carried away 
by all three inert states ($H$, $A$, $H^{\pm}$) can be counted 
as missing energy ($\slashed{E}_T$). 
In what follows, we assume both splittings are below $p_T^{\ell, {\rm min}}$ 
to show the maximal LHC sensitivity to the compressed limit through 
these searches. For our Monte Carlo analyses, we therefore set 
$m_H=m_A=m_{H^\pm}$ for simplicity.

Current ATLAS and CMS mono-jet searches using 8 TeV data do not constrain 
the compressed IDM for $m_H>m_Z/2$. We have verified this through 
a parton-level 
analysis  considering the limits on $\sigma\times \mathcal{A}\times \epsilon$ 
from the ATLAS search in Ref.~\cite{Aad:2015zva}, where $\mathcal{A}$ 
is the signal acceptance and $\epsilon$ the efficiency. Events are simulated 
in \texttt{MadGraph5}~\cite{Alwall:2014hca} and analyzed using the selection 
criteria outlined in Ref.~\cite{Aad:2015zva}, assuming $\epsilon=1$ for
the signal. Considering the nine signal regions defined 
in Ref.~\cite{Aad:2015zva}, we find $\sigma\times \mathcal{A}\times \epsilon/\left(\sigma\times \mathcal{A}\times \epsilon\right)_{95\% {\rm CL_s}}<0.5$ 
for all values of $m_H>m_Z/2$. Although this analysis is done at parton level, 
we do not expect showering or detector effects to make a qualitatively 
significant difference, due to the various vetoes on additional objects 
besides the hard jet and missing energy. We have explicitly verified 
this to be the case for the 14 TeV projections discussed below.

 %%%%%%%%
\begin{figure}[ttt]
\begin{center}
\includegraphics[width = 0.55\textwidth]{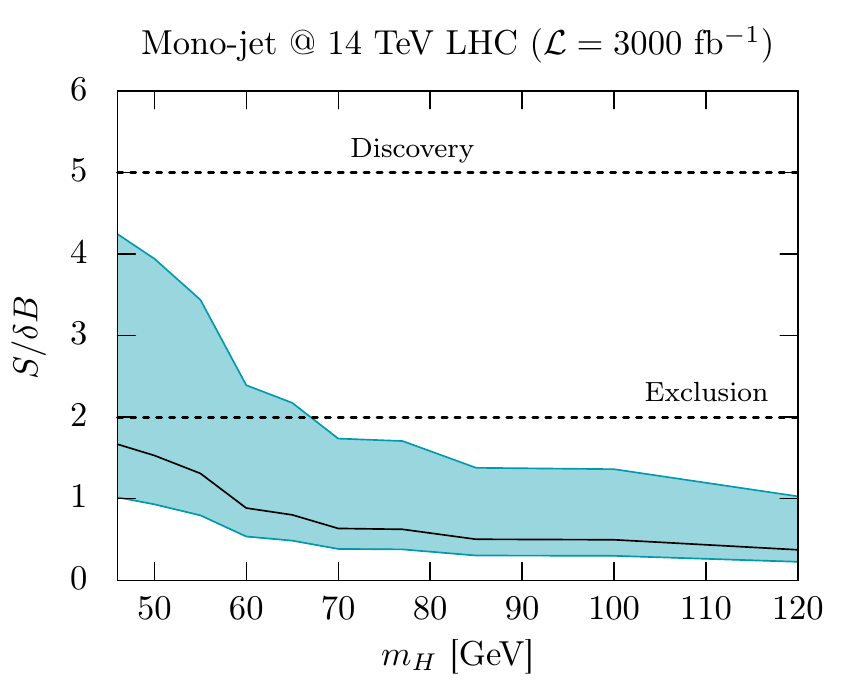}
\end{center}
\caption{\small  Predicted signal significance in the mono-jet channel at the 14 TeV LHC with integrated luminosity $\mathcal{L}=3000$ fb$^{-1}$ as a function of $m_H$ for the specific set of cuts described in Appendix~\ref{app}. 
The solid black line corresponds to a background systematic uncertainty 
$\epsilon_{\rm bkg}$ of 3\%. The shaded band was obtained by 
varying $\epsilon_{\rm bkg}$ from $1\!-\!5\%$. 
}
\label{fig:14tev}
\end{figure}
%%%%%%%%

Mono-jet searches at the 14 TeV~LHC may begin to test the compressed IDM 
with sufficiently high integrated luminosity provided systematic uncertainties 
on the backgrounds can be tightly controlled.  To estimate the sensitivity
of these searches, signal and background events are generated in 
\texttt{Madgraph5}~\cite{Alwall:2014hca}, 
showered in \texttt{Pythia6}~\cite{Sjostrand:2006za}, and processed
in \texttt{Delphes3}~\cite{deFavereau:2013fsa} for fast detector simulation.
We then impose the cuts and selection criteria described in Appendix~\ref{app}.
The results of this analysis are shown in in Fig.~\ref{fig:14tev}, 
where we plot the predicted mono-jet signal significance at 14~TeV
with $\mathcal{L}=3000\;\text{fb}^{-1}$ of data as a function of the mass $m_H$.
The significance shown in this figure is defined to be
\begin{equation}
\frac{S}{\delta B} \equiv \frac{S}{\sqrt{B + \epsilon_{\rm bkg}^2 B^2+ \epsilon_{\rm sig}^2 S^2}} \ ,
\end{equation}
where $S$ is the signal, $B$ is the background, and $\epsilon_{\rm sig}$
and $\epsilon_{\rm bkg}$ are their uncertainties.  

Our simulations were performed at leading order including up to two
hard jets at the parton level. We do not expect a full next-to-leading order (NLO)
calculation to significantly alter our results. The largest background contributions are from
$p p \rightarrow Z j (+X) \rightarrow \nu \overline{\nu} j (+X)$, and $p p \rightarrow W^{\pm} j (+X) \rightarrow \ell \nu j (+X)$, which involve the same 
diagrams as the dominant signal contributions with the neutrinos and charged leptons replacing the
inert scalars in the final state (see Eqs.~\ref{eq:ha}--\ref{eq:hpm}). 
We therefore expect NLO corrections to affect the signal and background distributions similarly. 
Moreover, the most significant NLO QCD corrections to the background are from 
the emission of an additional hard jet~\cite{Denner:2012ts}, which we account for by considering up to one 
additional jet at the parton level. Our minimum $p_T$ cut on the second jet eliminates
collinear enhancements in the cross-section, and so we expect virtual NLO effects 
to be relatively small. 

The shaded region 
in Fig.~\ref{fig:14tev} is the result of varying the systematic uncertainty 
$\epsilon_{\rm bkg}$ in the range $[0.01, 0.05]$, 
with the solid black line corresponding to $\epsilon_{\rm bkg}=0.03$. This range 
corresponds to that suggested by ATLAS for 14 TeV LHC mono-jet searches in Ref.~\cite{ATLAS_note}. The  
background systematics are dominated by uncertainties in the absolute jet energy scale and resolution, 
in Monte Carlo simulations, and in the determination of lepton momenta. It is difficult to predict the evolution 
of systematic uncertainties over the course of an experiment. However, we expect the range shown to approximate
both the near- and long-term prospects for the 14 TeV LHC, in line with the ATLAS analysis in Ref.~\cite{ATLAS_note}.  
The signal systematic uncertainty was set to $\epsilon_{\rm sig}=0.1$, 
following the analysis of Ref.~\cite{Low:2014cba} for compressed Higgsinos.  

All points considered in Fig.~\ref{fig:14tev} feature at least 10 signal events.  We also perform
an analogous parton-level analysis and find that it reproduces the full 
result for $S/\delta B$ within 10-15\% across the entire mass range considered. 

The results shown in Fig.~\ref{fig:14tev} suggest that the 14 TeV LHC 
can probe masses up to about 70 GeV, provided background systematic 
uncertainties are controlled to within 1\%. Such small uncertainties 
are likely unrealistic~\cite{Baer:2014cua,Han:2013usa}, although they may be possible
in the long-term~\cite{ATLAS_note}. 
If the background systematics are larger than about 2\%, 
this search will not be able to exclude any of the $m_H>m_Z/2$ region 
at the 14 TeV LHC. It is possible that a harder 
missing energy cut could provide slightly more sensitivity.  
However, the background systematics are likely to increase substantially 
as one moves farther out on the high-$\slashed{E}_T$ tail. 

  Some improvement in sensitivity over standard mono-jet searches might be
obtained by demanding additional soft leptons in the event from $H^\pm$ 
or $A$ decays.  This technique has shown promise for probing compressed 
electroweakinos~\cite{Low:2014cba, Giudice:2010wb,Gori:2013ala,Schwaller:2013baa, Han:2014kaa,Baer:2014kya,Han:2015lma} and 
sleptons~\cite{Rolbiecki:2012gn,Barr:2015eva}
in supersymmetric models.  However, comparing the compressed IDM to
scenarios with degenerate Higgsinos, this strategy does not appear to help
very much.  In the IDM, we are primarily interested in small $\Delta^0$
and larger $\Delta^{\pm}$.  For Higgsinos, the asymptotic decoupling limit 
is $\Delta^0 \simeq 2\Delta^{\pm}$~\cite{Han:2014kaa,Han:2015lma}.  
The most promising compressed Higgsino searches therefore focus on 
$\chi_2^0\to \chi_1^0Z^*$ decays with $Z^*\to \ell\ell$ since the leptons 
are likely to be more energetic, but also because the correlated 
kinematics between them allows for a good discrimination from background.
In the compressed IDM with small $\Delta^0$, the observable leptons will come
primarily from $H^+\to HW^*$ with $W^*\to\ell\nu_{\ell}$,
and so are less likely to pass the soft-lepton kinematic selection criteria 
in these searches~\cite{Han:2014kaa,Han:2015lma}.
For very small neutral mass splittings the $A\to HZ^*$ decay can be displaced, 
and searches for a mono-jet with a soft displaced vertex are also 
possible, as suggested for models of inelastic DM 
in Refs.~\cite{Bai:2011jg,Weiner:2012cb,Izaguirre:2015zva}.
In the IDM, we find that macroscopic decay lengths $c\tau > 1\,\text{mm}$
only occur for $\Delta^0\lesssim 1\,\gev$, 
and thus the displaced decay products of the $A\to HZ^*$ decays 
are typically too soft to be identified by the LHC detectors.

Taken together, our results suggest that the LHC will be unlikely 
to probe the compressed IDM much above $m_Z/2$, except perhaps in the 
mono-jet channel at 14 TeV provided provided background systematics can
be greatly suppressed.  For this reason, we turn next to investigate
the sensitivity of future colliders to this scenario.

\section{Future Colliders\label{sec:fc}}

To conclude our study, we consider the prospects for testing the compressed IDM at a future 100 TeV $pp$ collider and an ILC-type $e^+e^-$ collider.

\subsection{100 TeV $pp$ Collider}

A 100 TeV proton-proton collider is likely to significantly surpass the current LHC sensitivity to electroweakinos~\cite{Low:2014cba,Acharya:2014pua,Gori:2014oua,Bramante:2014tba,Badziak:2015qca}, and so it is worthwhile to consider its possible impact on the compressed IDM. To do so, we use \texttt{Madgraph5}~\cite{Alwall:2014hca} to perform a parton-level analysis of searches with a hard jet and large missing energy, as described in Appendix~\ref{app}. We expect that including showering and detector effects will not significantly alter the results. This was true of our 14 TeV LHC mono-jet study. Note also that no detector design currently exists for a 100 TeV collider.

The results of our analysis are shown in Fig.~\ref{fig:Fig8} for a representative set of cuts and selection criteria detailed in Appendix~\ref{app}.  The shaded region in this figure corresponds to varying the systematic uncertainty on the background $\epsilon_{\rm bkg}$ between $1\%\!-\!5\%$.  As expected, the reach is substantially higher than that of the 14 TeV LHC. However, masses above around 200 GeV will likely be difficult to probe. Comparing these results to those of Ref.~\cite{Low:2014cba}, which studied compressed Higgsinos at a 100 TeV collider, the reach we find is about a factor of four lower in mass due to the smaller production cross-section in the IDM. 

Mono-jet searches requiring additional soft leptons or disappearing tracks were also shown to be promising for compressed Higgsinos in Ref.~\cite{Low:2014cba}.  The former is unlikely to significantly improve on the pure mono-jet search in the compressed IDM for the reasons discussed in Sec.~\ref{sec:monojet}. Meanwhile, disappearing track searches are very sensitive to the chargino lifetime, which is governed by the chargino-neutralino mass splitting.  For a pure Higgsino, this splitting is about $\Delta^{\pm} \simeq 355\,\mev$~\cite{Thomas:1998wy}.  Since we generally expect larger values of $\Delta^{\pm}$ in the IDM, these disappearing track searches will not typically be applicable. We therefore expect that Fig.~\ref{fig:Fig8} represents a reasonable sensitivity estimate for a future 100 TeV $pp$ collider, given the information currently available about such a facility.

 %%%%%%%%
\begin{figure}[ttt]
\begin{center}
\includegraphics[width = 0.55\textwidth]{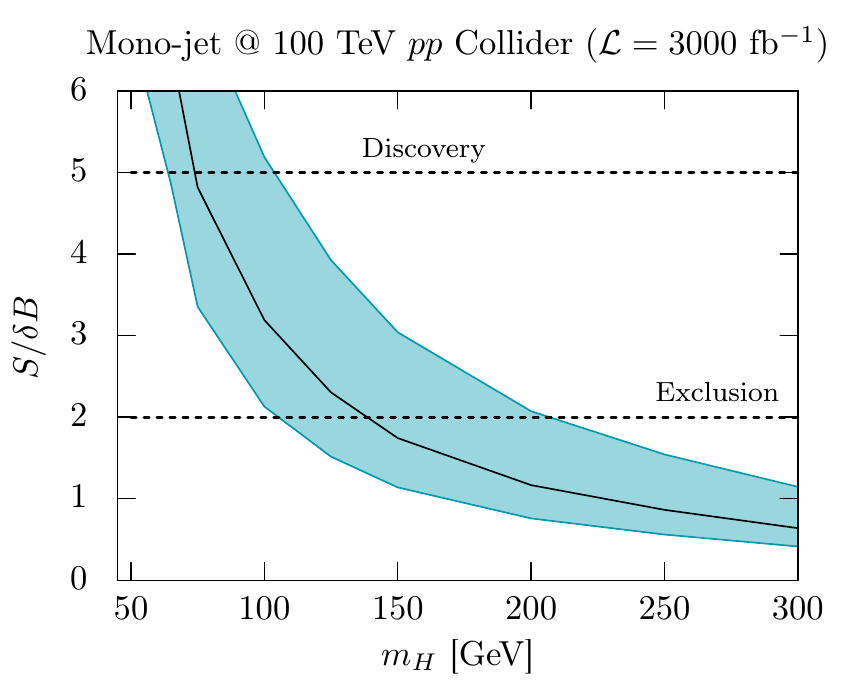}
\end{center}
\caption{\small  Estimated signal significance in the mono-jet channel at a 100 TeV $pp$ collider as a function of $m_H$ for a particular set of cuts, detailed in Appendix~\ref{app}. The black line corresponds to a background systematic error $\epsilon_{\rm bkg}=3\%$, while the shaded band was obtained by varying $\epsilon_{\rm bkg}$ from $1\!-\!5\%$. 
}
\label{fig:Fig8}
\end{figure}
%%%%%%%%

\subsection{High-energy Lepton Collider}

A high-energy $e^+e^-$ ILC-type collider would provide an additional tool to probe the compressed IDM.  The clean environment of such a machine, together with the option for polarized beams and low-threshold detectors, is well-suited for this scenario.

To estimate the reach of a potential ILC for the compressed IDM, 
we reinterpret the analysis of Ref.~\cite{Choi:2015zka}
which studied a light ``slepton'' scenario consisting of a $(2,-1/2)$ 
electroweak scalar doublet $(\widetilde{\nu},\,\widetilde{\ell})$ 
with only direct couplings to the electroweak vector bosons of the SM. 
A tree-level ($D$-term) mass splitting is assumed to push 
the charged state to be slightly heavier such that it decays promptly 
to the lighter state through an off-shell $W$ boson.   
The analysis of Ref.~\cite{Choi:2015zka} demands a single hard
photon and large missing energy, and finds an exclusion ($2\sigma$)
reach of about $140\,\gev$ ($160\,\gev$) 
for $\gamma\widetilde{\nu}\widetilde{\nu}^*$
($\gamma\widetilde{\ell}\widetilde{\ell}^*$) 
production with $500\,\text{fb}^{-1}$ of data 
at $\sqrt{s}=500\,\gev$ and making use of polarized beams
to reduce the dominant background from $e^+e^-\to \gamma\nu\bar{\nu}$.
Considerably stronger bounds are found for light degenerate 
Higgsinos due once again to the lower net production rates for scalars 
and the faster turn off of their production 
cross-sections as one approaches the kinematic threshold. 

The ``slepton'' model studied in Ref.~\cite{Choi:2015zka} can
be mapped directly onto the compressed IDM with 
$\widetilde{\nu} \leftrightarrow (H+iA)/\sqrt{2}$ 
and $\widetilde{\ell}\leftrightarrow H^-$,
which have the same gauge quantum numbers.  The limits of this analysis
are directly applicable to $HA$ and $H^+H^-$ production in the very compressed
limit of $\Delta^{0,\pm} \lesssim 1\,\gev$.  However, since no attempt is made 
to distinguish between the charged ($\widetilde{\ell}$) 
and neutral ($\widetilde{\nu}$) states, and no veto on additional objects 
is imposed, we expect that these bounds will not be degraded 
too severely for larger mass splittings up to 
the roughly $\Delta \lesssim 10\,\gev$ we are most interested in.  

  Based on the results of our LEP analyses, the ILC sensitivity to the
compressed IDM with $\Delta^{\pm}\gtrsim 3\,\gev$ can likely 
be improved further relative to the pure monophoton analysis
of Ref.~\cite{Choi:2015zka} by searching for soft leptons in addition
to a hard photon and missing energy.  This strategy has been applied
to nearly-degenerate Higgsinos in the ILC studies of 
Refs.~\cite{Berggren:2013vfa,Baer:2013vqa,Berggren:2013bua,Baer:2014yta}.
Since an ILC-type detector is expected to have a very low lepton $p_T$ threshold, 
perhaps approaching $0.5\,\gev$~\cite{Behnke:2013lya}, it might even
be possible to improve the sensitivity for even smaller mass splittings.
It would be interesting to revisit this possibility once the sensitivities 
of an ILC detector are better known.  For larger mass splittings
$\Delta^{\pm}\gtrsim 10\,\gev$, the mono-photon requirement can be dropped
and searches can be performed for leptons (or jets) 
and missing energy as described in Refs.~\cite{Asano:2011aj,Aoki:2013lhm}.

\section{Conclusions\label{sec:conc}}

In this study, we have performed a detailed analysis of the inert doublet model
in the compressed limit. Mass degeneracies can arise in the presence of
approximate continuous symmetries. The compressed scenario allows for light
electroweakly--charged scalars that are compatible with electroweak precision
observables, results from dark matter experiments, and cosmological
constraints. Unfortunately, these new states can be difficult to test at
colliders.

We have carefully analyzed the constraints from LEP on the compressed IDM. A
reinterpretation of the OPAL dilepton plus MET search excludes masses up to
about 90 GeV for charged-neutral mass splittings above 5 GeV. However, there
are currently no direct constraints on masses above $m_Z/2$ 
for charged-neutral mass splittings less than 5 GeV. 
These conclusions are a refinement on past work,
which had previously focused on mass splittings above several GeV.

Current 8 TeV LHC results yield no additional constraints beyond LEP in the
compressed region. Leptons are typically too soft to trigger on in exclusive
lepton searches, and the mono-jet production cross-sections are significantly
smaller than those for Higgsinos. However, the 14 TeV LHC may begin to probe low
masses in the mono-jet channel, although this depends quite sensitively on the
systematic uncertainties for the backgrounds.  Our results suggest that
uncertainties below $\sim 2-3\%$ may be sufficient to exclude masses only slightly
above $m_Z/2$ with 3000 fb$^{-1}$ of data. We also do not expect searches for
mono-jet topologies with additional soft decay products, charged tracks, or
displaced vertices to significantly alter this conclusion. 

Future colliders can offer considerably better opportunities for testing the
compressed IDM. A 100 TeV $pp$ may be able to probe inert doublet masses
up to 100--200~GeV in the monojet channel, with the precise limit depending
on the degree to which systematic uncertainties can be controlled 
(with a range of 1--5\% assumed here).  For an ILC-like $e^+ e^-$ collider
with center-of-mass energy $\sqrt{s} = 500\,\gev$ and $500\,\text{fb}^{-1}$
of data, we find a sensitivity up to masses of $m=140\,\gev$ ($160\,\gev$)
for $HA$ ($H^+H^-$) in the highly compressed limit using a pure mono-photon
search.  An even larger reach is expected away from the compressed limit
by including leptons in the search channels. Ultimately, however, these future
experiments are unlikely to probe the region of the compressed IDM in which the
lightest inert scalar saturates the observed dark matter abundance via thermal
freeze-out. In this high-mass region, direct detection efforts are likely to
offer the most promise. Nevertheless, a combination of LHC, ILC, and 100 TeV
collider searches would be valuable in conclusively testing low- to
intermediate-mass scalars in the compressed inert doublet model.

\section*{Acknowledgments}

  We thank Eder Izaguirre, Robert McPherson, Brian Shuve, and Isabel Trigger
for helpful discussions.  This work is supported by the Natural Sciences
and Engineering Research Council of Canada~(NSERC).

\appendix

\section{Mono-jet Projections}\label{app}

In this appendix, we detail our mono-jet analysis of the compressed IDM.

\subsection{14 TeV LHC}

\begin{figure}[ttt]
\begin{center}
\includegraphics[width = 0.47\textwidth]{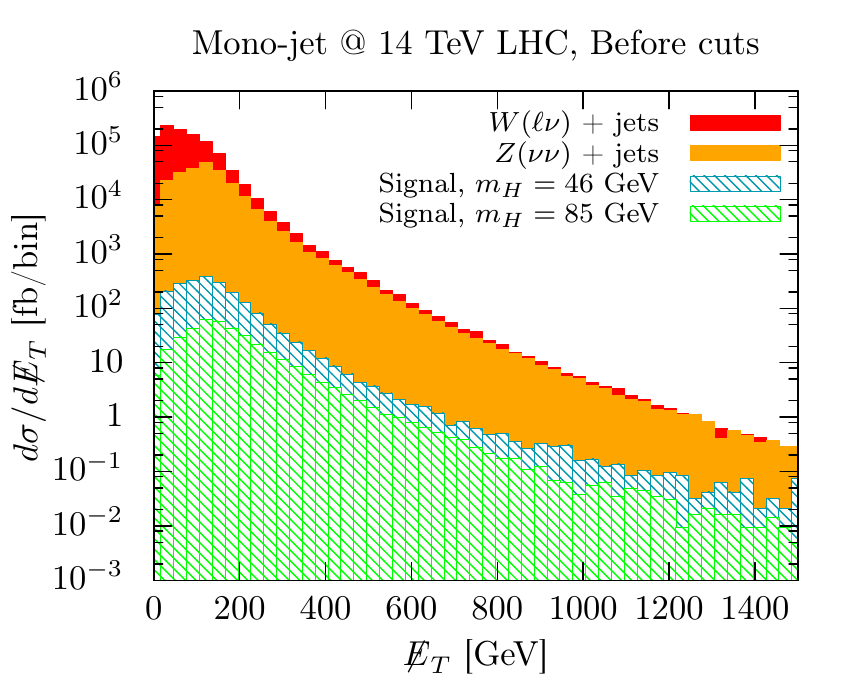}\hspace{0.1cm}\includegraphics[width = 0.47\textwidth]{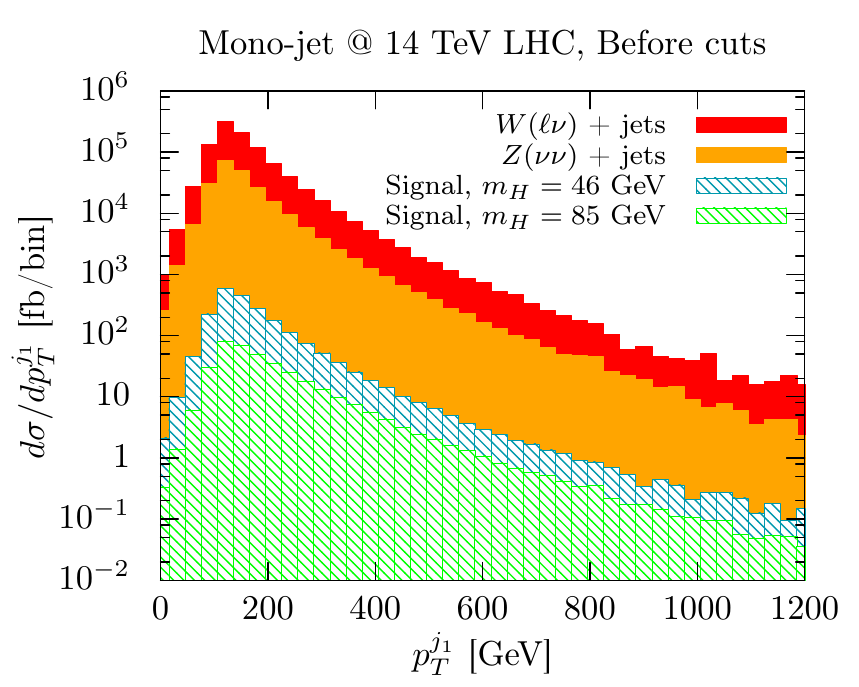}\\
\includegraphics[width = 0.47\textwidth]{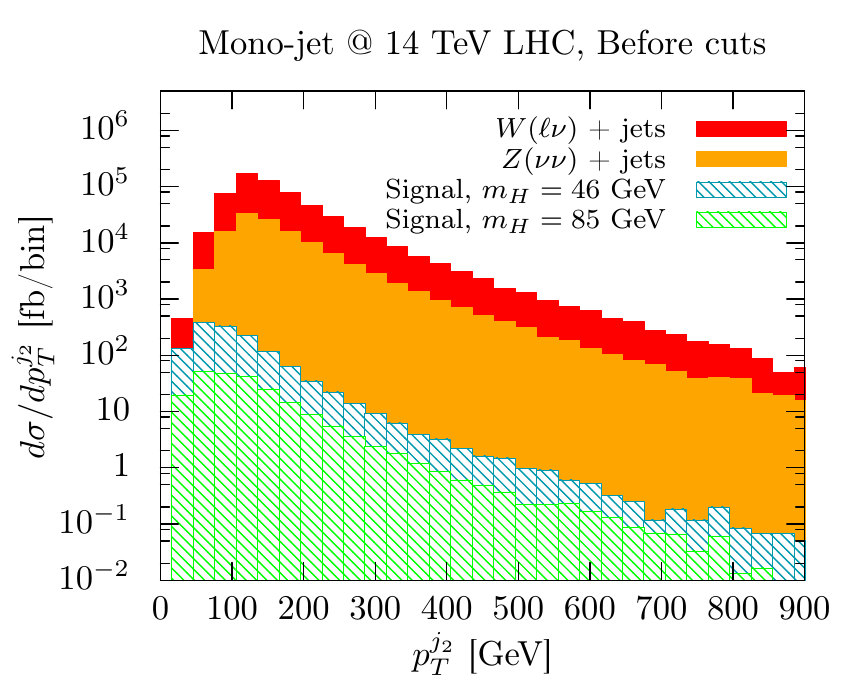}\hspace{0.1cm}\includegraphics[width = 0.47\textwidth]{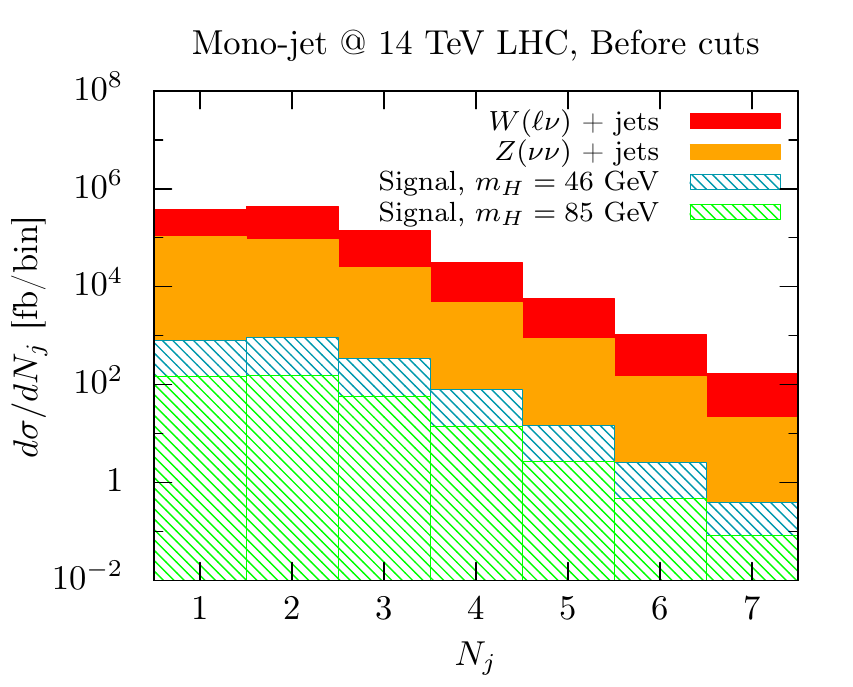}
\end{center}
\caption{\small Sample histograms of the signal and background events generated for our 14 TeV LHC mono-jet analysis before cuts. The shapes of all the signal distributions mimic those of the background, making discrimination from the background difficult. Note that, after cuts, the $Z$ background dominates the signal regions we consider.}
\label{fig:Fig9}
\end{figure}

We perform our 14 TeV LHC analysis at the detector level, using \texttt{MadGraph5}~\cite{Alwall:2014hca} to generate events,  \texttt{Pythia6}~\cite{Sjostrand:2006za} to shower and hadronize them, and \texttt{Delphes3}~\cite{deFavereau:2013fsa} as a fast detector simulator. Our strategy is similar to that found in previous mono-jet studies~\cite{Low:2014cba,Han:2013usa,Baer:2014kya,ATLAS_note}. Specifically, we select events if they satisfy the following criteria:
\begin{equation}
\begin{aligned}
&p_T^{j_1}>300 \, {\rm GeV}, \, \, \left|\eta_{j_1}\right|<2.0, \, \, \slashed{E}_T>1 \, {\rm TeV}, \\ \Delta&\phi(j_{1,2},\slashed{E}_T)>0.5, \, \, p_T^{j_2}<100\,{\rm GeV,} \, \, N_{j}\leq 2
\end{aligned}
\end{equation}
where $j_1$ is the jet with the largest $p_T$, and identified jets are required to have $p_T>50$ GeV and $\left|\eta\right| <3.6$. Note that these jet definitions were suggested by ATLAS in Ref.~\cite{ATLAS_note} for studies of 14 TeV mono-jet searches. Up to one additional identified jet is allowed; however we require $p_T^{j_2}<100$ GeV, since the background $p_T^{j_2}$ distribution peaks at slightly higher values than the signal for low $m_H$ (see the bottom left panel of Fig.~\ref{fig:Fig9}). We also include the following lepton vetoes:
\begin{itemize}
\item Veto on $e$ with $p_T(e)>7$ GeV, $\left|\eta(e)\right|<2.5$
\item Veto on $\mu$ with $p_T(\mu)>7$ GeV, $\left|\eta(\mu)\right|<2.47$
\item Veto on hadronic taus, with  $p_T(\tau_h)>20$ GeV, $\left|\eta(\tau_h)\right|<2.3$.
\end{itemize}
We use the default CMS detector card implemented in \texttt{Delphes3} and neglect pile-up effects. In addition to the cuts shown above, we consider several different values of $p_T^{j_1,{\rm min}}$, $p_T^{j_2,{\rm min}}$ as suggested in Refs.~\cite{Low:2014cba,Han:2013usa, Baer:2014kya} for Higgsinos, and we vary $\slashed{E}_T^{\rm min} \in$ [400, 1400] GeV. The final choices listed represent one of the more promising configurations.  Tighter MET cuts may offer slightly more sensitivity, however we expect the uncertainties in both the background and signal to increase quickly as $\slashed{E}_T^{\rm min}$ is raised.

The dominant backgrounds are $Z$ + jets, where $Z\rightarrow \overline{\nu} \nu$, and $W$+jets where the $W$ decays leptonically. We validate our backgrounds by checking against the results of Refs.~\cite{Han:2013usa, Baer:2014kya}. The distributions of the various background events in $\slashed{E}_T$ and $p_T^{j_1}$ before cuts are shown in Fig.~\ref{fig:Fig9}, along with the signal for two different masses. The distribution of signal events closely resembles that of the backgrounds for large $\slashed{E}_T$ and $p_T^{j_1}$, making discrimination difficult. Note that, after cuts, the $Z$+jets background dominates in the signal region defined above.

 %%%%%%%%
%%%%%%%%

\subsection{100 TeV $pp$ Collider}

We perform our 100 TeV analysis at parton level, since no detector design currently exists for such a collider. Signal and background events are generated in \texttt{Madgraph5}, and we neglect pile-up effects. The dominant backgrounds in this case are once again $Z$ + jets, where $Z\rightarrow \overline{\nu} \nu$, and $W$+jets where the $W$ decays leptonically. We once again neglect the effect of pile-up. Our selection criteria are similar to those used for our LHC analysis, with higher cuts on the various transverse momenta. Specifically, we require:
  \begin{equation}
\begin{aligned}
p_T^{j_1}>&1.2 \, {\rm TeV}, \, \, \left|\eta_{j_1}\right|<2.0, \, \, \slashed{E}_T>5 \, {\rm TeV}, \\ &\Delta\phi(j_{1,2},\slashed{E}_T)>0.5, \, \, N_{j}\leq 2
\end{aligned}
\end{equation}
where identified jets are required to have $p_T>250$ GeV and $\left|\eta\right| <3.6$. We do not include an upper limit on $p_T^{j_2}$, as we find it has little effect on the distributions in this case. Following Ref.~\cite{Low:2014cba}, the lepton vetoes are chosen as follows:
\begin{itemize}
\item Veto on $e$ with $p_T(e)>20$ GeV, $\left|\eta(e)\right|<2.5$
\item Veto on $\mu$ with $p_T(\mu)>20$ GeV, $\left|\eta(\mu)\right|<2.1$
\item Veto on hadronic taus, with  $p_T(\tau_h)>40$ GeV, $\left|\eta(\tau_h)\right|<2.3$.
\end{itemize}

We investigate the effect of varying the cut on missing transverse energy between 2.5--7.0 TeV. As in the LHC analysis, tighter MET cuts may slightly improve the reach, especially for cases with larger $\epsilon_{\rm bkg}$, however once again the systematic uncertainties in the signal and backgrounds are expected to increase significantly for larger $\slashed{E}_T^{\rm min}$.  Note that Ref.~\cite{Low:2014cba} used similar cuts for studying compressed Higgsinos at a 100 TeV collider.

%%%%%%%%%%%%%%%%%%%%%%%%%%%%%%%%%%%%%%%%%%%%%%%%%%%%%%%%%%%%%%%%%

\end{document}